\documentstyle[multicol,pre,aps,epsf]{revtex}
\newcommand{\ftriangle}[1]{%
\setlength{\unitlength}{#1} %{0.6mm}
\begin{picture}(5,3.2)%\thicklines
\put(4,0){\line(-1,0){4}}
\put(3.885,0.2){\line(-1,0){3.769}}\put(3.769,0.4){\line(-1,0){3.538}}
\put(3.654,0.6){\line(-1,0){3.307}}\put(3.538,0.8){\line(-1,0){3.076}}
\put(3.423,1.0){\line(-1,0){2.845}}\put(3.307,1.2){\line(-1,0){2.614}}
\put(3.192,1.4){\line(-1,0){2.383}}\put(3.076,1.6){\line(-1,0){2.152}}
\put(2.961,1.8){\line(-1,0){1.922}}\put(2.845,2.0){\line(-1,0){1.691}}
\put(2.730,2.2){\line(-1,0){1.460}}\put(2.614,2.4){\line(-1,0){1.229}}
\put(2.499,2.6){\line(-1,0){0.998}}\put(2.383,2.8){\line(-1,0){0.767}}
\put(2.326,2.9){\line(-1,0){0.651}}\put(2.268,3.0){\line(-1,0){0.534}}
\put(2.210,3.1){\line(-1,0){0.420}}
\end{picture}}

\newcommand{\fhako}[1]{%
\setlength{\unitlength}{#1} %{0.6mm}
\begin{picture}(5,3.2)\thicklines
\put(4,0){\line(-1,0){4}}
\put(4,0.2){\line(-1,0){4}}\put(4,0.4){\line(-1,0){4}}
\put(4,0.4){\line(-1,0){4}}\put(4,0.6){\line(-1,0){4}}
\put(4,0.8){\line(-1,0){4}}\put(4,1.0){\line(-1,0){4}}
\put(4,1.2){\line(-1,0){4}}\put(4,1.4){\line(-1,0){4}}
\put(4,1.6){\line(-1,0){4}}\put(4,1.8){\line(-1,0){4}}
\put(4,2.0){\line(-1,0){4}}\put(4,2.2){\line(-1,0){4}}
\put(4,2.4){\line(-1,0){4}}\put(4,2.6){\line(-1,0){4}}
\put(4,2.8){\line(-1,0){4}}\put(4,3.2){\line(-1,0){4}}
\put(4,3.2){\line(-1,0){4}}\put(4,3.4){\line(-1,0){4}}
\put(4,3.6){\line(-1,0){4}}\put(4,3.8){\line(-1,0){4}}
\put(4,4.0){\line(-1,0){4}}
\end{picture}
}

\begin{document}
\draft
\title{Glass transitions and dynamics in thin polymer films:\\
dielectric relaxation of thin films of polystyrene
}
\author{K. Fukao\cite{A} and Y. Miyamoto}
\address{
Faculty of Integrated Human Studies,
Kyoto University, Kyoto 606-8501,
Japan 
}
\date{Received \today}
\maketitle
\begin{abstract}
The glass transition temperature $T_{\rm g}$ and the temperature 
$T_{\alpha}$ corresponding to the peak in the dielectric loss 
due to the $\alpha$-process
have been simultaneously determined as functions of film thickness~$d$ 
through dielectric measurements for polystyrene thin films supported on 
glass substrate. The dielectric loss peaks have also been investigated as  
functions of frequency for a given temperature. 
A decrease in $T_{\rm g}$ was observed with decreasing film thickness,
while $T_{\alpha}$ was found to remain almost constant for $d>d_{\rm c}$ and 
to decrease drastically with decreasing $d$ for $d<d_{\rm c}$. Here,
$d_{\rm c}$ is a critical thickness dependent on molecular weight. 
The relaxation time $\tau_{\alpha}$ of the $\alpha$-process, which 
was measured as the frequency at which the dielectric loss realizes its peak 
value at a given temperature, was found to have a $d$ dependence similar 
to that of $T_{\alpha}$.
The relaxation function for the $\alpha$-process was obtained by using 
the observed frequency dependence of the peak profile of the dielectric loss.  
The exponent $\beta_{\mbox{\tiny KWW}}$, which was obtained 
from the relaxation functions, decreases as thickness decreases. 
This suggests that the distribution of relaxation times for the 
$\alpha$-process broadens with decreasing thickness.
The thickness dependence of $T_{\rm g}$ is directly related to the distribution
of relaxation times for the $\alpha$-process, not to the relaxation time itself. 
The value of the thermal expansion coefficient normal to the film
surface was found to increase with decreasing film thickness below
$T_{\rm g}$, but to decrease with decreasing film thickness above $T_{\rm g}$.  
These experimental results are discussed in the context of a three-layer
model in which within thin films there are three layers with different 
mobilities and glass transition temperatures. 

\end{abstract}
%64.70.Pf  Glass transitions
%68.60.-p  Physical properties of thin films, nonelectronic
%77.22.Gm  Dielectric loss and relaxation

%\vspace{-0.3cm}
\begin{multicols}{2}
\section{Introduction}
Recently, glass transitions in amorphous materials 
have been investigated by many researchers. 
However, the mechanism involved in  glass transitions is not yet fully 
understood~\cite{Noncry,Pisa,Ediger1}. Understanding the behavior of the 
characteristic length scale of 
the dynamics of supercooled liquids near the glass transition 
is the most important problem to be solved in such studies.
In Adam and Gibbs' theory, it is assumed that there is a domain 
in which collective particle motion can occur and its 
size grows as the temperature is lowered.
This domain is called the {\it cooperatively rearranging region} 
(CRR)~\cite{Adam-Gibbs}. In connection with the CRR, recent
molecular dynamics simulations have revealed the existence of significant 
large-scale heterogeneity in particle displacements, so-called
{\it dynamical heterogeneity} in 
supercooled liquids~\cite{Muranaka,Perera,Yamamoto1,Kob}. As the 
temperature decreases toward 
$T_{\rm g}$, the dynamical heterogeneity grows. 
Experimental studies using multi-dimensional NMR~\cite{Spiess}, 
dielectric hole burning~\cite{Bohmer} and photobleaching~\cite{Ediger2}
have produced evidence of dynamical heterogeneity.
These topics concerning heterogeneity are closely  
related to the length scale of dynamics near glass transitions.

Glass transitions in finite systems confined to
nanopores~\cite{Schuller,Kremer1,Barut} and thin 
films~\cite{Keddie1,Keddie2,DeMaggio,Forrest1,Forrest1a,Wallace,Jerome1,Jerome2} 
have recently attracted much attention, because such systems can be regarded 
as model systems for studying the length scale of glass transitions.  
In such systems, deviation from  
bulk properties is expected to appear if the system size is comparable
to the characteristic length scale. In particular, $T_{\rm g}$
and the thermal expansion coefficient $\alpha_{\rm n}$ of thin films have  
been measured using several experimental techniques, including 
ellipsometry~\cite{Keddie1,Keddie2}, positron annihilation lifetime 
spectroscopy (PALS)~\cite{DeMaggio}, 
Brillouin light scattering~\cite{Forrest1,Forrest1a}, and X-ray 
reflectivity~\cite{Wallace}. 
For the first time, Keddie {\it et al.} investigated $T_{\rm g}$ and the 
thermal expansion coefficient $\alpha_{\rm n}$ of thin polymer films supported on
substrate. For polystyrene films on hydrogen-passivated Si, $T_{\rm g}$ 
was found to decrease with decreasing film thickness $d$ for 
$d$$<$40nm~\cite{Keddie1}. The value of $\alpha_{\rm n}$ below $T_{\rm g}$ 
was found to increase with decreasing $d$ approaching the value
characterizing liquid states. It was suggested that this decrease in
$T_{\rm g}$ is caused by the presence of a liquid-like layer at the  
polymer-air interface in this case; 
in the case of freely standing polystyrene films, $T_{\rm g}$ decreases 
{\it much more rapidly} with decreasing film thickness~\cite{Forrest1}. 
These results suggest that the
interaction between polymers and the substrate competes with surface
effects. This competition leads to a more gradual decrease of $T_{\rm g}$ 
in the former case. For a strong 
attractive interaction between polymers and the substrate, an increase in 
$T_{\rm g}$ with decreasing $d$ was observed~\cite{Keddie2}. 

Positron annihilation lifetime measurements reveal that the observed 
$T_{\rm g}$ values of supported PS films are similar to 
those obtained by Keddie {\it et al}. However,
the thermal expansion coefficient obtained by PALS is independent 
of $d$ below $T_{\rm g}$, while it decreases with decreasing thickness 
above $T_{\rm g}$. 
It was proposed that there is a dead layer near the interface 
between polymers and the substrate in addition to a liquid-like layer 
at the polymer-air interface. In the case of thin polymer films supported on 
substrate, the glass transition temperature and thermal properties 
strongly depend on the competetion between interfacial and surface 
effects. There are still controversial experimental results for such systems.

The dynamics related to the glass transition in thin films have
been investigated using several methods~\cite{Hall,Forrest2,Kajiyama}.   
Second harmonic generation reveals that the distribution of relaxation
times broadens with decreasing film thickness, while the average
relaxation time of the $\alpha$-process remains constant for 
supported films of a random copolymer~\cite{Hall}. Ultrasonic 
measurements~\cite{Forrest2} have shown that the temperature $T_{\rm max}$ at 
which the ultrasonic absorption exhibits a maximum for a given frequency 
has a $d$ dependence 
similar to that of $T_{\rm g}$ obtained by ellipsometric measurements 
for thin polystyrene films supported on substrate. 
In the case of freely standing films of polystyrne, photon correlation 
spectroscopy~\cite{Forrest2} studies indicate that the relaxation behavior of 
the $\alpha$-process in thin films is similar to that of bulk samples 
of polystyrene, except for the reduction of the $\alpha$-relaxtion time.
Atomic force microscopy studies~\cite{Kajiyama} have revealed the 
existence of a mobile layer near the free surface of films of 
polystyrene.  Because there are only a few experimental observations on 
the dynamics of thin polymer films, it is not yet clear whether properties of 
the $\alpha$-process change together with $T_{\rm g}$ 
as the film thickness decreases or whether the obtained results 
depend on the methods used for measurements or on the details of the
individual samples.

Dielectric measurement is one of the most powerful expermental tools to 
investigate the dynamics of the $\alpha$-process in amorphous materials.
Recently we applied this method to the determination of the glass transition 
temperature through measurements of the thermal expansion coefficient~\cite{Fukao1}.
Bauer {\it et al.} also used this method and further extended it to 
thermal expansion spectroscopy~\cite{Bauer1,Bauer2}.   
By virture of dielectric measurements, it is possible to simultaneously measure the 
glass transition temperatures and determine the relaxation behavior of the 
$\alpha$-process of {\it a single sample} even for thin films.  

In a previous paper~\cite{Fukao1}, we reported that $T_{\rm g}$ for thin
polystyrene films supported on glass substrate can be determined from
the temperature change of the electric capacitance during heating and 
cooling processes 
and that the dynamics of the $\alpha$-process can be determined from the  
dielectric loss of the films. We were able to obtain the distinct thickness 
dependences of $T_{\rm g}$ and $T_{\alpha}$ in which the dielectric loss 
exhibits a peak value for a fixed frequency due to the $\alpha$-process.
In this paper, the results obtained through dielectric measurements are 
described in detail, and the dynamics of the $\alpha$-process are 
investigated by measurements of the frequecncy dispersion of the 
dielectirc loss for the purpose of clarifying the relation involving  
the thickness dependence of $T_{\rm g}$, the thermal 
expasion coefficients, and the dynamics of the $\alpha$-process.
Based on the results of {\it simultaneous} measurements, we discuss 
the relationship between $T_{\rm g}$ and the dynamics of the
$\alpha$-process of thin polymer films. 
A possible explanation of our experimental results for $T_{\rm g}(d)$ and the 
dynamics of the $\alpha$-process is given in terms of a three-layer model. 
This paper consists of five sections. In Sec.II, experimental details 
and principles regarding our determination of $T_{\rm g}$ and the thermal expansion 
coefficient using electric capacitance measurements are given. 
The experimental results for the thermal expansion coefficient 
$\alpha_{\rm n}$ and $T_{\rm g}$ obtained from our measurements
are given in Sec.III. A three-layer model, which 
can account for the observed thickness dependences of $\alpha_{\rm n}$ and 
$T_{\rm g}$, is also introduced there. In Sec.IV the dynamics of the
$\alpha$-process of thin films are investigated in reference to the peak 
profile in the dielectirc loss due to the $\alpha$-process in the frequency
domain. In Sec.V overall discussion and a summary of this paper are given. 
 
\section{Experimental details}
\subsection{Sample preparation and measurement procedures }
Four different atactic polystyrenes (a-PS) were used. These were 
purchased from Scientific 
Polymer Products, Inc. ($M_{\rm w}$=2.8$\times$10$^5$), the Aldrich 
Co., Ltd. ($M_{\rm w}$=
1.8$\times$10$^6$, $M_{\rm w}/M_{\rm n}$=1.03), and 
Polymer Source, Inc. ($M_{\rm w}$=
3.6$\times$10$^4$, $M_{\rm w}/M_{\rm n}$=1.06 and $M_{\rm w}$=
3.6$\times$10$^3$, $M_{\rm w}/M_{\rm n}$=1.06).
Thin films of a-PS with various thicknesses 
from 6 nm to 489 nm were prepared on an Al-deposited slide glass using a 
spin-coat method from a toluene solution of a-PS. 
The thickness was controlled by changing the concentration of the solution. 
After annealing at 70$^{\circ}$C in the vacuum system for several days 
to remove solvents, Al was vacuum-deposited again to 
serve as an upper electrode. Heating cycles in which the temperature was 
changed between room temperature 
and 110$^{\circ}$C ($>$$T_{\rm g}$) were applied prior to  
the dielectric measurements to relax the as-spun films and obtain 
reproducible results. 
`Bulk' films of a-PS (each with $d>$100$\mu$m) were made by oil-pressing  
samples melted at about 200$^{\circ}$C for a few minutes, and gold was 
vacuum-deposited onto both sides of the films to serve as electrodes.   
Dielectric measurements were done using an LCR 
meter (HP4284A) in the frequency range from 20 Hz to 1MHz during 
heating (cooling) processes in which the temperature was changed at a 
rate of 2K/min.
For dielectric measurements of very thin films, the resistance of 
the electrodes cannot be neglected.  This leads to an extra loss peak
on the high frequency side, which results from the fact that system is 
equivalent to a series circuit of a capacitor and resistor, where 
the capacitance is that of the sample and the resistance is that of the 
electrodes~\cite{Kremer2}. 
The peak shape in the frequency domain can be fitted well by a simple Debye-type 
equation.  Data obtained in the frequency domain, therefore, can be accurately 
corrected by subtracting the `C-R peak' and assuming the validity of the
Debeye equation.
Data corrected in this manner were used for further analysis in this paper.

\subsection{Relation involving the electric capacitance, thickness and thermal 
expansion coefficients}
In this section, we give the relation between the temperature change of the 
electric capacitance and thermal expansion coefficient. 
Similar discussion was also given by Bauer {\it et al.}~\cite{Bauer1}. 
In our measurements,
film thickness was evaluated from the capacitance at room temperature
of as-prepared films by using the formula for the capacitance $C'$ 
of a flat-plate condenser, $C'$=$\epsilon'\epsilon_0S/d$, where 
$\epsilon'$ is the permittivity of a-PS, $\epsilon_0$ is the permittivity 
of the vacuum, $S$ is the area of the electrode ($S$=8.0mm$^2$), and $d$ 
is the thickness of the 
films. 
In general, the geometrical capacitance is given by
\begin{eqnarray}\label{C0}
C_0(T)\equiv\epsilon_0\frac{S}{d}\sim\epsilon_0\frac{S_0}{d_0}
(1+(2\alpha_t-\alpha_n)\Delta T),
\end{eqnarray}
and the permittivity is expressed by
\begin{eqnarray}\label{eps1}
\epsilon'(\omega,T)=\epsilon_{\infty}(T)+\epsilon_{\rm disp}(\omega, T)
%\sum_{j}\Re
%\left[\frac{\Delta\epsilon_j(T)}{[1+(i\omega_j\tau_j(T))^{1-\alpha_j}]^{\beta_%j}}\right]
,
\end{eqnarray}
where $\epsilon_{\infty}$ is the permittivity in the high-frequency
limit, $\alpha_t$ is the linear thermal expansion coefficient parallel
to the film
surface, $\alpha_n$ is the linear thermal expansion coefficient normal to 
the film surface, $\Delta T=T-T_0$, and $T_0$ is a standard temperature. 
The second term $\epsilon_{\rm disp}$ on the right-hand side of 
Eq.(\ref{eps1}) is related to the frequency dispersion of the dielectric loss 
due to the $\alpha$-process, the $\beta$-process, and so on. 
%$\alpha_i$ and $\beta_j$ are shape parameters of the 
%Havriliak-Negami equation for the j-th process.

If we here assume that the films are constrained along 
the substrate surface, %the values of 
$\alpha_t$ and $\alpha_n$ are given by
\begin{eqnarray}\label{alpha_tn}
\alpha_t = 0\qquad \mbox{and}\qquad\alpha_n &=& \frac{1+\nu}{1-\nu}\alpha_{\infty},
\end{eqnarray}
where $\nu$ is Poisson's ratio and  $\alpha_{\infty}$ is the bulk linear 
coefficient of thermal expansion~\cite{Wallace}.
It should be noted that this case corresponds to that of `constant area
conditions' in Ref.~\cite{Bauer1}.

In the temperature range where the effect of the dielectric dispersion,
$i.e.$, the second term in the expression for $\epsilon'$, can be
neglected, we obtain
\begin{eqnarray}\label{eps1_infty}
\epsilon'(T)=\epsilon_{\infty}(T)\sim\epsilon_{\infty}(T_0)
(1-\eta_0\alpha_n\Delta T),
\end{eqnarray}
where 
\begin{eqnarray}
\epsilon_{\infty}(T_0)&=&\frac{1+2\xi_0}{1-\xi_0}\\
\eta_0&\equiv&\frac{3\xi_0}{(1-\xi_0)(1+2\xi_0)},\\
\xi_0&\equiv&\frac{1}{3\epsilon_0}\sum_j N_{j,0}\bar{\alpha_j}.
\end{eqnarray}
Here $N_{j,0}\equiv N_j(T_0)$, $N_j(T)$ is the number density of the
j-th atom at $T$, and $\bar{\alpha_j}$ 
is the polarizability of the j-th atom. 
The Clausius-Mossotti relation,
$(\epsilon_{\infty}-1)/(\epsilon_{\infty}+2)=1/3\epsilon_0\cdot\sum 
N_j\bar{\alpha_j}$,
where $N_j(T)=N_j(T_0)(1-\alpha_{\rm n}\Delta T)$, has been used.

In the case of a-PS, the dielectric constant $\epsilon '$ of bulk samples is 
2.8 at room temperature~\cite{Yano1}. If we assume that $\epsilon'(T_0)
\approx \epsilon_{\infty}(T_0)$=2.8, then $\xi_0=0.375$ and $\eta_0$ 
is nearly equal to 1. Using Eqs.(\ref{C0}), (\ref{alpha_tn}) and 
(\ref{eps1_infty}), we obtain the temperature 
coefficient of the capacitance $\tilde\alpha$ as follows:
\begin{eqnarray}\label{t_alpha0}
\tilde\alpha &\equiv& -\frac{1}{C'(T_0)}\frac{dC'(T)}{dT}\nonumber \\
&=& -\left(\frac{1}{\epsilon'(T_0)}\frac{d\epsilon'}{dT}+
\frac{1}{C_0(T_0)}\frac{dC_0}{dT}\right) \nonumber \\
&=&(1+\eta_0)\alpha_{\rm n}
\approx 2\alpha_n.
\end{eqnarray}
We thus see that the temperature coefficient of $C'$ is 
proportional to $\alpha_{\rm n}$.  
It is therefore expected that the temperature 
coefficient changes at $T_{\rm g}$. 
In the literature~\cite{PolymerHand} we find 
$\nu$=0.325 and $\alpha_{\infty}$=0.57$\times10^{-4}$/K for $T<T_{\rm g}$, 
and $\nu$=0.5 and $\alpha_{\infty}$=1.7$\times10^{-4}$/K for $T>T_{\rm g}$.
Hence, for bulk samples of a-PS, it can be expected that
\begin{eqnarray}\label{t_alpha}
\tilde\alpha =\left\{ \begin{array}{r@{\quad:\quad}l} 
2.2\times 10^{-4} & T < T_{\rm g} \\
10.2\times 10^{-4} & T > T_{\rm g}.
\end{array}\right.
\end{eqnarray}

\section{Glass transition temperature of thin films}
\subsection{Glass transition temperature and thermal expansion coefficients}
Figure 1 displays the temperature dependence of the capacitance, normalized 
with respect to the value at 303K during the heating processes. 
In Fig.1(a) we can see that the values at thickness 91 nm 
for different frequencies fall along a single line 
and decrease with increasing temperature for the temperature range 
from room temperature to approximately 370K.
At higher temperature the normalized capacitance decreases with increasing 
temperature more steeply than at lower
temperature. In this range, the values for different frequencies can no longer be 
fitted by a single line. Here, they are dispersed due to the appearance of the 
$\alpha$-process. For the temperature range shown in the
figure, however, it is apparent that the effect of the dispersion is
quite weak above 10kHz. Therefore, for such frequencies the temperature
at which the slope of $C'(T)$ changes discontinuously can be determined 
unambiguously as the crossover temperature between the
line characterizing the lower temperature side and that 
characterizing the higher
temperature side. This crossover temperature
can be regarded as $T_{\rm g}$, because the thermal expansion
coefficient changes 
%%%%%%%%%%% FIG.1 %%%%%%%%%%%
%\hspace*{-0.2cm}
\begin{minipage}{8.5cm}
\begin{figure}
\epsfxsize=7.9cm %\epsfysize=5cm
%\vspace{4cm}
\centerline{
%
%\vspace{3cm}
%\vspace*{-1.2cm}
\epsfbox{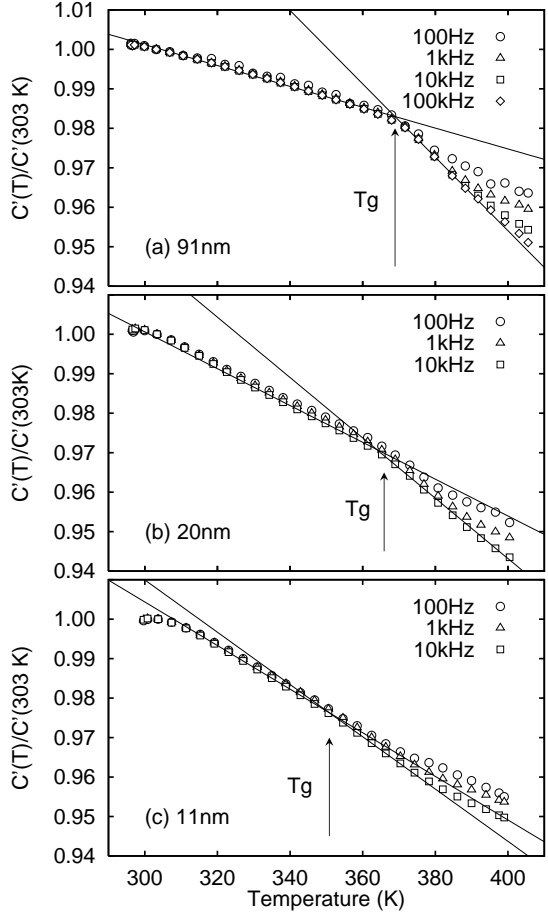}}
\vspace{-4.5cm}
\caption{Temperature dependence of the capacitance normalized with
respect to the
values at 303K during the heating process for various frequencies 
from 100Hz to 10kHz
%($\circ$ corresponds to 100Hz, $\triangle$ to 1kHz, $\Box$ to 10kHz, 
%and $\diamond$ to 100kHz) 
and three different thicknesses: 
((a) $d$=$91$nm and 
$M_{\rm w}$=1.8$\times$10$^6$; (b) $d$=$20$nm and $M_{\rm w}$=
2.8$\times$10$^5$; (c) $d$=$11$nm and $M_{\rm w}$=1.8$\times$10$^6$). 
The solid lines were obtained by fitting the data for 10kHz to a linear 
function below and above $T_{\rm g}$. The arrows 
indicate the values of $T_{\rm g}$.
}
\label{fig:fig1}
\end{figure}
\vspace{0.1cm}
%\vspace{-0.5cm}
\end{minipage}
%%%%%%%%%%%%%%%%%%%%%%%%%%%%%
through the crossover temperature, as can be 
expected from the discussion in Sec.II. 
%The temperature coefficients of $C'$  
%obtained by fitting the 
%data to the two lines are 2.6$\times$10$^{-4}$K$^{-1}$ 
%for $T<T_{\rm g}$ and 8.7$\times$10$^{-4}$K$^{-1}$ for $T>T_{\rm
%g}$ in the case $d$=91nm. These values agree well with those 
%estimated from the values of Eq.(\ref{t_alpha}) calculated from 
%$\alpha_{\infty}$ and $\nu$ in the literature.
As $d$ decreases, $T_{\rm g}$ also decreases, as shown in Figs.1(b) and (c). 

Figure 2 displays the $d$ dependence of $T_{\rm g}$ for a-PS films with 
four different molecular weights. In each case, the values of $T_{\rm g}$  
for the various values of $d$ were determined as the crossover temperatures 
at which the temperature 
coefficient of the capacitance at 10 kHz changes during 
heating process.  For thick films, the values of $T_{\rm g}$ are 
almost equal to those for bulk PS. When the films are 
thinner than about 100 nm, however, a decrease in $T_{\rm g}$ is observed. 
The value of $T_{\rm g}$ for films of 6 nm thickness is lower by 
about 30K than 
that of films of 489 nm thickness for a-PS films with $M_{\rm w}$=
2.8$\times$10$^5$. The dependence of $T_{\rm g}$ on $d$ can be expressed 
as 
\begin{eqnarray}\label{Tg_d} 
T_{\rm g}(d)=T_{\rm g}^{\infty} \left(1-\frac{a}{d}\right),
\end{eqnarray}
%%%%%%%%%%% FIG.2 %%%%%%%%%%%
%\hspace*{-0.2cm}
%\vspace*{-1cm}
\begin{minipage}{8.5cm}
\begin{figure}
\epsfxsize=9cm %\epsfysize=5cm
%\vspace*{8cm}
\centerline{
\hspace*{-0.6cm}\epsfbox{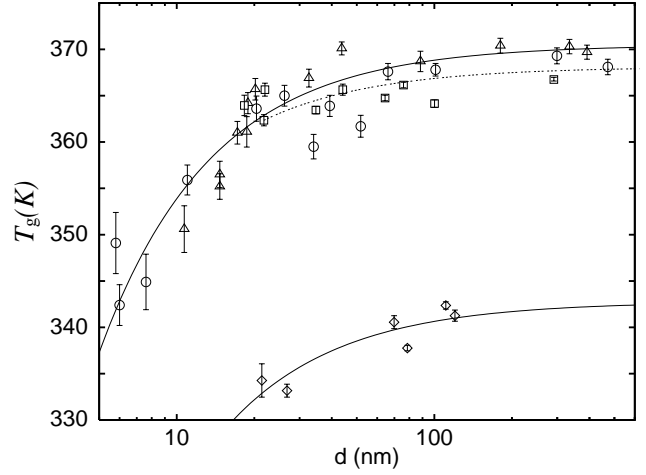}
}
\vspace{0.4cm}
\caption{Thickness dependence of $T_{\rm g}$ 
of a-PS films obtained during the heating process ($\triangle$
corresponds to $M_{\rm w}$=1.8$\times$10$^6$, 
$\circ$ to $M_{\rm w}$=2.8$\times$10$^5$, 
$\Box$ to $M_{\rm w}$=3.6$\times$10$^4$, and $\Diamond$ to 
$M_{\rm w}$=3.6$\times$10$^3$). 
The values of $T_{\rm g}$ are determined as the crossover temperatures 
between the lines characterizing $C'(T)$ at 10kHz below and 
above $T_{\rm g}$. The curves were obtained from the equation 
$T_{\rm g}(d)=T_{\rm g}^{\infty}(1-a/d)$. 
The upper solid curve is for $M_{\rm w}$=1.8$\times$10$^6$ and 
2.8$\times$10$^5$, the lower solid curve is for $M_{\rm w}$=
3.6$\times$10$^3$, and the dotted curve is for $M_{\rm w}$=
3.6$\times$10$^4$.
}
\label{fig:fig2}
\end{figure}

\vspace{0.0cm}
\end{minipage}
%%%%%%%%%%%%%%%%%%%%%%%%%%%%%
%%%%%%%%%%%%%% Table I %%%%%%%%%%%%%
%\hspace*{-0.2cm}
\begin{minipage}{8.5cm}
\begin{table}
%\vspace*{5cm}
\caption{The values of the parameters resulting in the best fit of 
the glass transition temperatures $T_{\rm g}(d)$ to Eq.(\protect\ref{Tg_d}) 
for thin films of a-PS with various molecular weights 
($M_{\rm w}$=3.6$\times$10$^3$$\sim$ 1.8$\times$10$^6$).}
\begin{tabular}{cccc} % In second brace, l = left, r = right,
% c = centered and d = decimal justification.
$M_{\rm w}$ & $a$ (nm) & $T_{\rm g}^{\infty}$ (K)\\\tableline 
1.8$\times$10$^6$, 2.8$\times$10$^5$ & 0.45$\pm$0.06 & 370.5$\pm$0.9\\
3.6$\times$10$^4$ & 0.33$\pm$0.05 & 367.8$\pm$0.5\\
3.6$\times$10$^3$ & 0.62$\pm$0.12 & 342.8$\pm$1.1
% One&Two\tablenote{footnote.}&Three&Four\\ % Place \tablenote{}
\end{tabular}
\end{table}
\end{minipage}
%%%%%%%%%%%%%%%%%%%%%%%%%%%%%%%%%%%%
where $T_{\rm g}(d)$ is the measured glass transition temperature for a
film of thickness $d$. The values of the parameters resulting in the
best fit are listed in Table I. 

The asymptotic value $T_{\rm g}^{\infty}$ has a distinct molecular weight 
dependence. For bulk samples of a-PS, the variation in $T_{\rm g}$ with
molecular weight is described well by the empirical equation 
\begin{eqnarray}\label{Fox}
T_{\rm g}^{\infty}=\tilde T_{\rm g}^{\infty}-\frac{C}{N},
\end{eqnarray}
where $N$ is the degree of polymerization, $\tilde T_{\rm g}^{\infty}$=373 
K and C=1.1$\times$10$^3$~\cite{Fox1}. Using Eq.(\ref{Fox}), we obtain 
the values of $T_{\rm g}^{\infty}$ as follows: $T_{\rm g}^{\infty}$= 373 K for 
$M_{\rm w}$=1.8$\times$10$^6$ and 2.8$\times$10$^5$, 370 K for $M_{\rm w}$
=3.6$\times$10$^4$, and 341 K for $M_{\rm w}$=3.6$\times$10$^3$. 
Therefore, it is found that the $M_{\rm w}$-dependence of $T_{\rm g}$ for 
the bulk samples can be reproduced quite well by the present measurements.

%%%%%%%%%%% FIG.3 %%%%%%%%%%%
\hspace*{-0.35cm}
\begin{minipage}{8.5cm}
\begin{figure}
\epsfxsize=8.5cm %\epsfysize=5cm
%\vspace*{8cm}
\centerline{
\hspace*{0cm}\epsfbox{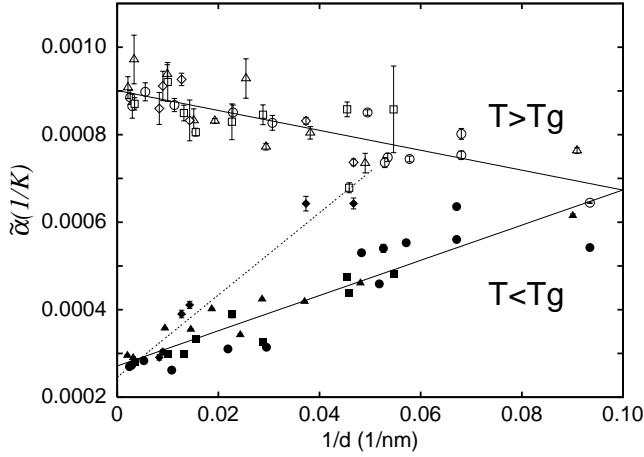}
}
\vspace{0.4cm}
\caption{Thickness dependence of the temperature coefficient of the capacitance of 
a-PS films obtained by measurements of the electric capacitance at 10kHz 
during the heating process. Above $T_{\rm g}$, 
the symbol $\protect\circ$ correspond to $M_{\rm w}$=1.8$\times$10$^6$, 
$\triangle$ to $M_{\rm w}$=2.8$\times$10$^5$, $\Box$ to 
$M_{\rm w}$=3.6$\times$10$^4$, and $\Diamond$ to 
$M_{\rm w}$=3.6$\times$10$^3$. Below $T_{\rm g}$, the filled symbol 
are used, and the correspondence between the symbols and the molecular 
weights is the same as above $T_{\rm g}$.  
%$\bullet$ correspond to $M_{\rm w}$=1.8$\times$10$^6$, 
%\protect\ftriangle{0.6mm} to 
%$M_{\rm w}$=2.8$\times$10$^5$,\protect\fhako{0.5mm}to 
%$M_{\rm w}$=3.6$\times$10$^4$, and \protect\fdiamond{0.6mm} to 
%$M_{\rm w}$=3.6$\times$10$^3$.
The lines were obtained by fitting the data to 
Eqs.(\protect\ref{alpha1}) and (\protect\ref{alpha2}).
The dotted line is valid only 
for the data of the films with $M_{\rm w}$=3.6$\times$10$^3$ below $T_{\rm g}$.
}
\label{fig:fig3}
\end{figure}

\vspace{0.1cm}
%\vspace{-0.3cm}
\end{minipage}
%%%%%%%%%%%%%%%%%%%%%%%%%%%%%

As shown in Table I, the length scale $a$ related to the thickness
dependence of $T_{\rm g}$ ranges from 0.3-0.6nm. Taking into account 
the scatter in $a$, however, there is no distinct molecular weight
dependence of $a$. This length scale is of the same order as that of the statistical 
segment of polystyrene (0.68nm)~\cite{Ballard1}.
These experimental results for $T_{\rm g}^{\infty}$ and 
$a$ suggest that 
the thickness dependence of $T_{\rm g}$ is almost independent of the molecular 
weight of a-PS after rescaling with respect to $T_{\rm g}$ of bulk samples.
This seems to be consistent with the fact that the length scale 
$a$ is not related to chain lengths but rather to segment lengths.

%$T_{\rm g}^{\infty}=370.2\pm 0.4$~K and $a=0.43\pm0.03$~nm. 
If our data are fitted by the function proposed by Keddie {\it et al.}, 
$T_{\rm g}(d)=T_{\rm g}^{\infty}\left(1-\left(a/d %frac{a}{d}
\right)^{\tilde\delta}\right)$, 
the observed data for 
$M_{\rm w}$=1.8$\times$10$^6$ and 2.8$\times$10$^5$ can also be fitted
to this equation, and the parameter values resulting in the best fit
are as follows: $a=0.39\pm 0.10$nm and $\tilde\delta=0.96\pm 0.08$.
The revised value of $\tilde\delta$ obtained by Keddie and Jones is $\tilde\delta
=1.28\pm 0.20$~\cite{Keddie3}.  This  suggests that Keddie's equation
can be replaced by Eq.(\ref{Tg_d})~($\tilde\delta\approx$1). 
The values $T_{\rm g}$ obtained in our measurements also agree well with
those obtained by Forrest {\it et al.}~\cite{Forrest1a}. 
%and also those by Keddie et al, after rescaling the data with respect 
%to $T_{\rm g}$ of the bulk samples. 
Thus, it can be concluded that $T_{\rm g}$ of thin films has
been accurately determined by measurements of the electric capacitance.
With the present method it has been confirmed that  the apparent 
$T_{\rm g}$ obtained in our measurements decreases with decreasing film thickness. 
 
Since no general theory for the effect on the glass transition 
of finite size has yet been presented, Eq.(\ref{Tg_d}) is just an experimental
findings at the present stage. As discussed later, in the context of the 
three-layer model we regard 
the glass transition temperature described by Eq.(\ref{Tg_d}) as that 
associated with the $\alpha$-process (in this case, segmental motions 
of polymer chains) in the 
{\it bulk-like layer} between the dead layer and the liquid-like one. 
The observed decrease in $T_{\rm g}$ may be attributed to the
distribution of the relaxation times of the $\alpha$-process, which
comes from the existence of boundary between liquid-like layer and
bulk-like layer, and also between bulk-like layer and dead layer.

Here, it
should be noted that the temperature coefficient $\tilde\alpha$ of
electric capacitance also changes with film thickness, as shown in Fig.1. 
To clarify the $d$ dependence of $\tilde\alpha$, Fig.3 displays
$\tilde\alpha$ as a function of the inverse of the film thickness for thin
films. 
It is found that $\tilde\alpha$ increases with decreasing
thickness below $T_{\rm g}$, while it decreases with decreasing film 
thickness above $T_{\rm g}$. In both cases, the thickness dependence of 
$\tilde\alpha$ can be expressed as {\it a linear function of the 
inverse of film thickness}.
This observed $d$ dependence of $\tilde\alpha$ seems to be 
independent of the molecular weight within experimental accuracy,
except for the case with $M_{\rm w}$=3.6$\times$10$^3$ below $T_{\rm g}$. 
The values of $\tilde\alpha$ for bulk samples can be obtained by taking $1/d$
to zero as follows: $\tilde\alpha$=9.0$\times$10$^{-4}$ K$^{-1}$
for $T>T_{\rm g}$ and 2.8$\times$10$^{-4}$ K$^{-1}$ for $T<T_{\rm g}$.
These values agree well with those predicted by Eq.(\ref{t_alpha}) 
in Sec.II, and hence 
the temperature coefficient of the electric capacitance $\tilde\alpha$ 
observed in the present measurements can be regarded as the 
linear thermal expansion coefficient normal to the substrate 
$\alpha_{\rm n}$ multiplied by a factor of 2, as shown 
in Eq.(\ref{t_alpha0}).
This $d$ dependence of $\tilde\alpha$ and $\alpha_{\rm n}$ can be 
attributed to  
the existence of some layers with different chain mobilities within 
thin polymer films supported 
on substrate, as discussed in the next section.

\vspace*{0.3cm}\subsection{Three-layer model}
In order to explain the observed thickness dependence of $\tilde\alpha$, 
we introduce a {\it three-layer model}, in which it is assumed that a thin
polymer film on substrate consists of three layers with different
molecular mobilities~\cite{DeMaggio}. Near the interface between the 
glass substrate and
polymers there is a {\it dead layer} which has almost no mobility. 
On the other hand, near the free surface there is a {\it liquid-like layer} 
which has higher mobility. 
Here, we assume that the thicknesses of the two layers are $\delta$ 
(dead layer) and $\xi$ (liquid-like layer). 
Between these two layers there is a {\it bulk-like 
layer} which has the same mobility as that of the bulk samples. 
In this model, below the {\it apparent} $T_{\rm g}$ the observed linear thermal
expansion coefficient $\alpha_{\rm n}$ normal to the surface of the
substrate is given by
\begin{eqnarray}\label{alpha1}
\alpha_{\rm n} = \frac{\xi}{d}\alpha_{\rm l}^{\infty}+\left(1-\frac{\delta+\xi}
{d}\right)\alpha_{\rm g}^{\infty},
\end{eqnarray}
and above $T_{\rm g}$ by
\begin{eqnarray}\label{alpha2}
\alpha_{\rm n} = \left(1-\frac{\delta}{d}\right)\alpha_{\rm l}^{\infty},
\end{eqnarray}
where $\alpha_{\rm g}^{\infty}$ ($\alpha_{\rm l}^{\infty}$) is the
linear thermal expansion coefficient of the bulk glassy (liquid) state.
Therefore, this simplified model can reproduce the observed thickness 
dependence of $\tilde\alpha$(=2$\alpha_{\rm n}$) both below and above $T_{\rm g}$.
By fitting the observed results given in Fig.3 to Eqs. (\ref{alpha1}) and
(\ref{alpha2}), the thicknesses of dead and liquid-like layers are 
obtained: $\delta$=$(2.5\pm 0.3)$nm and $\xi$=$(7.5\pm 0.3)$nm.
%These $d$-dependences of $\tilde\alpha$ have also been observed by other 
%methods such as ellipsometry~\cite{Keddie1} and positron 
%annihilation~\cite{DeMaggio}. 
Keddie {\it et al.} estimated the thickness of a
liquid-like layer near the free surface to be (8.0$\pm$0.8)nm.  
DeMaggio {\it et al.} obtained the thickness of the dead layer between polymer 
films and substrate to be (5.0$\pm$0.5)nm and also proposed the
existence of a mobile surface layer with thickness 2 nm. 
The values of $\delta$ and $\xi$ observed in the present dielectric measurements 
are found to be consistent with those obtained by other experimental
techniques: PALS and ellipsometry.  
Therefore, we conclude that the three layer model can successfully be
applied in our case.

The values of $\delta$ and $\xi$ for the lowest molecular weight sample 
($M_{\rm w}$=3.6$\times$10$^3$) are $\delta$=(3.7$\pm$1.0)nm and 
$\xi$=(15.2$\pm$0.9)nm.  
These values are clearly different from those for other molecular
weights. This difference may be due to the entanglement effect, 
because the critical molecular weight at the entanglement limit 
is 1.3$\times$10$^4$ for a-PS~\cite{Phys1}.
In the present measurements no obvious molecular weight dependence
was observed within experimental accuracy except for 
$M_{\rm w}$=3.6$\times$10$^3$. However, it is plausible that the values 
of $\delta$ and $\xi$ 
are functions of the molecular weight. More precise measurements will be 
required to detect such molecular weight dependence.

Thin polymer films with both substrate and upper electrodes of Al for 
dielectric measurements were used in the present measurements.  Due to 
the presence of these electrodes, for our thin films there
was no real free surface existing in the air-polymer interface.
The samples discussed presently were  {\it capped} supported films, according to the 
terminology given in Ref.\cite{Forrest1a}. However, the values of $T_{\rm g}$ 
observed in the present measurements
agree with those obtained by ellipsometry for uncapped films. 
It has also been reported that there is no obvious differences between 
the results obtained for uncapped and capped supported
films~\cite{Forrest1a}. 
These experimental results support the conclusion that there is no appreciable 
difference in $T_{\rm g}$ for capped and uncapped supported thin
films; {\it i.e.}, 
the upper electrode of our samples can be assumed to have no effect on the 
thermal properties of the polymer films, in particular the thermal 
expansivity along the direction normal to the substrate. 
Recent DSC measurements have revealed the existence of a surface mobile layer 
of PS spheres dispersed in Al$_2$O$_3$ powders~\cite{Tasaka1}.
According to the discussion given by Mayes~\cite{Mayes1}, the glass transition 
temperature of a (surface) layer can be depressed only if the end concentration
of the layer is higher than that of bulk samples. The existence of 
a true 
%%%%%%%%%%% FIG.4 %%%%%%%%%%%
\hspace*{-0.35cm}
\begin{minipage}{8.5cm}
%\vspace*{-0.5cm}
\begin{figure}
\epsfxsize=9.0cm %\epsfysize=5cm
%\vspace*{8cm}
\centerline{
%\vspace{-0.3cm}
\hspace*{-0.6cm}\epsfbox{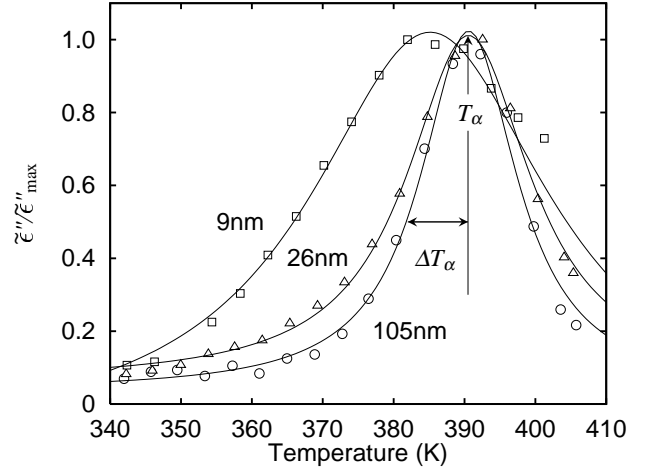}
}
\vspace{0.3cm}
\caption{Reduced dielectric loss as a function of temperature 
for various film thicknesses ($M_{\rm w}$=2.8$\times$10$^5$ and $f$=100Hz). 
The symbol $\protect\circ$ corresponds to $d=105$ nm, 
$\triangle$ to $d=26$ nm, and $\Box$ to $d=9$ nm. The curves were
obtained by fitting the data to 
the equation $\tilde\epsilon''=\tilde\epsilon''_{\rm max}/(1+((T-T_{\alpha })/\Delta T_{\alpha })^2)$.
}
\label{fig:fig4}
\end{figure}

%\vspace{0.1cm}
\vspace{-0.5cm}
\end{minipage}
%%%%%%%%%%%%%%%%%%%%%%%%%%%%%
free surface is not a necessary condition for the decrease of 
$T_{\rm g}$ and the existence of a liquid-like layer.

Recent measurements of the mass density of a-PS thin films supported on 
Si using  neutron 
reflectivity show that the average mass density within films is 
near the bulk value regardless of film thickness~\cite{Wallace2}.
Sound velocities in thin freely standing PS films measured  by Brillouin 
light scattering are reported to be the same for all films 
with various film thicknesses. This also suggests that the average 
mass density 
of thin films is the same as that of bulk samples~\cite{Forrest3}. 
In the presently considered layer model it is assumed that there are 
thin liquid-like and dead layers in addition to the layer with bulk properties.  
Because the liquid-like layer has a lower mass density and the dead
layer has a higher mass density than the bulk layer, it is not
unreasonable to assume that the average mass density of these thin films 
is the same as that of bulk samples. 
In the case of freely standing films, because there is no dead layer,
the average mass density is expected to become lower for very thin
films than that of bulk sample. However, the observed value of the 
average mass density does not change with film thickness. The simple 
layer model may no longer be valid for freely standing films, and 
it may be the case that 
another physical factor must be taken into account. 

It should also be noted here that picosecond acoustic techniques reveal
an increase in the longitudinal sound velocity for thin films of
poly(methyl methacrylate) and PS~\cite{Lee1,Bretz1}.  This suggests 
a change in the average mass density for thin films from that of  
bulk samples. This result disagrees with that obtained in 
Ref.\cite{Wallace2,Forrest3}.  

%%%%%%%%%%% FIG.5 %%%%%%%%%%%
\hspace*{-0.35cm}
%\vspace{-0.5cm}
\begin{minipage}{8.5cm}
\begin{figure}
\epsfxsize=7.8cm %\epsfysize=10cm
%\epsfxsize=7.8cm %\epsfysize=5cm
%\vspace*{5cm}
\centerline{
\epsfbox{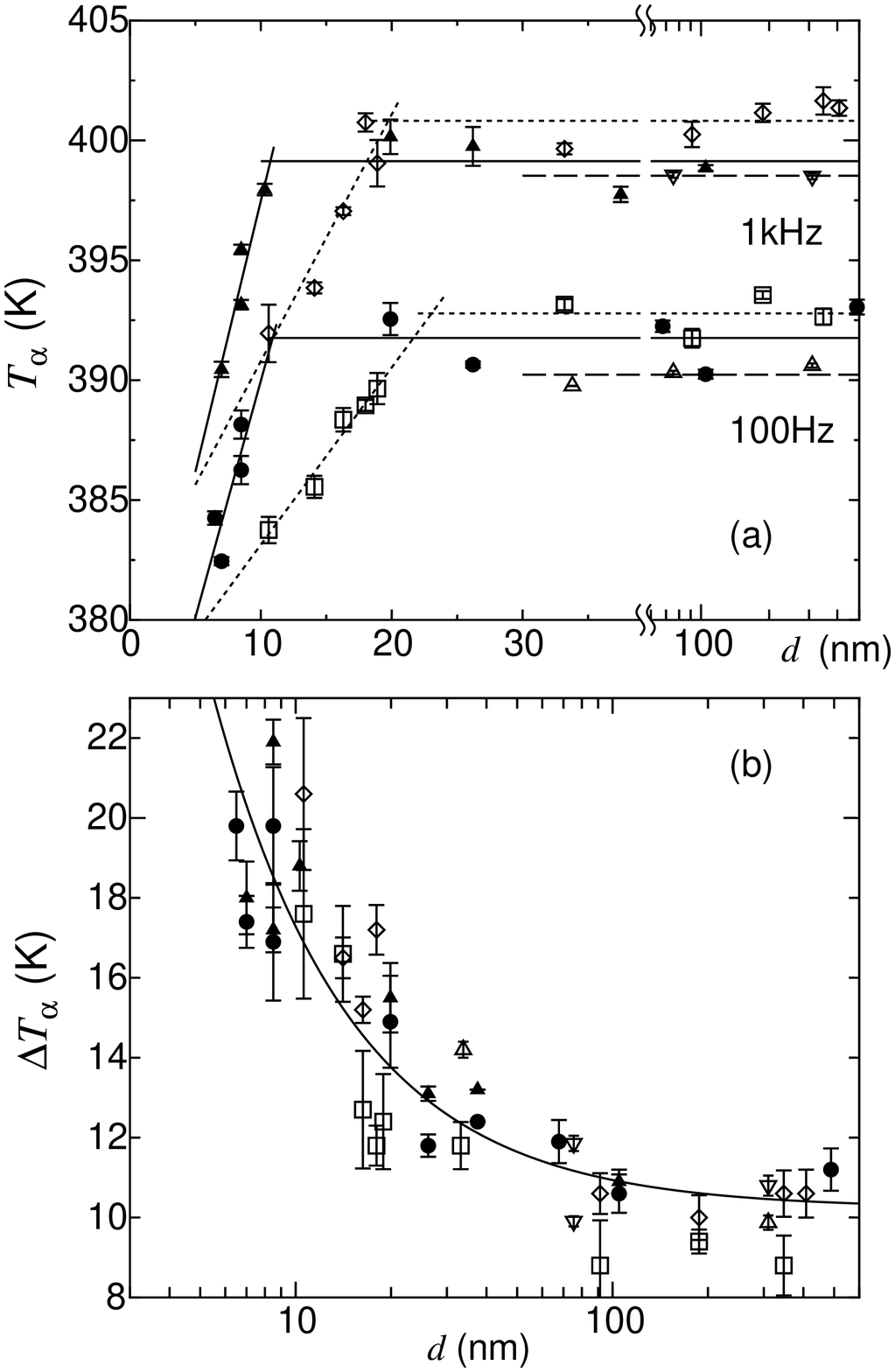}
}
\vspace{0.3cm}
\caption{Thickness dependence of (a) $T_{\alpha}$ and (b) $\Delta T_{\alpha}$
during the heating 
process for a given frequency. The symbol 
$\bullet$ corresponds to $f$=100Hz and $M_{\rm w}$=2.8$\times$10$^5$,% 
\protect\ftriangle{0.5mm} to $f$=1kHz and $M_{\rm w}$=2.8$\times$10$^5$, 
$\Box$ to $f$=100Hz and $M_{\rm w}$=1.8$\times$10$^6$,  
$\Diamond$ to $f$=1kHz and $M_{\rm w}$=1.8$\times$10$^6$, 
$\triangle$ to $f$=100Hz and $M_{\rm w}$=3.6$\times$10$^4$, and
$\bigtriangledown$ to $f$=1kHz and $M_{\rm w}$=3.6$\times$10$^4$. 
The solid lines in (a) were drawn for $M_{\rm w}$=2.8$\times$10$^5$, 
the dotted lines, for $M_{\rm w}$=1.8$\times$10$^6$, and 
the broken lines, for $M_{\rm w}$=3.6$\times$10$^4$. 
These lines were drawn using Eq.(\protect\ref{Ta_d}) and 
the curve in (b) was obtained from Eq.(\protect\ref{Del_alpha}).
}
\label{fig:fig5}
\end{figure}

\vspace{-0.4cm}
\end{minipage}
%%%%%%%%%%%%%%%%%%%%%%%%%%%%%

\section{Dynamics of the $\alpha$-process of thin films}
In this section, results concerning dielectric loss during the heating process 
are given to allow for discussion how the dynamics of the $\alpha$-process 
change with decreasing $T_{\rm g}$ as a result of decreasing thickness.
First, the temperature dependence of the dielectric loss with a fixed 
frequency is investigated to directly compare the thickness dependence 
of $T_{\rm g}$ with that of the dynamics of the $\alpha$-process. 
Second, the results for the dielectric loss in the frequency domain
under an isothermal condition are given to confirm the results 
obtained with fixed frequency and to allow for discussion of the relaxation behavior 
due to the $\alpha$-process in thin polymer films.

%%%%%%%%%%%%%% Table II %%%%%%%%%%%%%
\hspace*{-0.2cm}
\begin{minipage}{8.5cm}
\begin{table}
%\vspace*{5cm}
\caption{The values of $T_{\alpha}^{\infty}$, $d_{\rm c}$ 
and  $\zeta$ resulting in the best fit for 
thin films of a-PS with $M_{\rm w}$=2.8$\times$10$^5$ and 1.8$\times$10$^6$.}
\begin{tabular}{ccccc} % In second brace, l = left, r = right,
% c = centered and d = decimal justification.
$M_{\rm w}$ & $f$ (Hz) & $T_{\alpha}^{\infty}$(K) & $d_{\rm c}$ (nm) & 
$\zeta$ (10$^2$nm) \\\tableline 
2.8$\times$10$^5$ & 100 & 391.8$\pm$0.5 & 10.9$\pm$1.1 & 2.0$\pm$0.7\\     
                  & 1000 & 398.9$\pm$0.4 & 10.3$\pm$0.7 & 1.6$\pm$0.5\\   
1.8$\times$10$^6$ & 100 & 392.8$\pm$0.3 & 23.1$\pm$0.9 & 5.3$\pm$0.6\\   
                & 1000 & 400.8$\pm$0.4 & 19.7$\pm$0.8 & 3.9$\pm$0.5\\    
%33                & 0.38$\pm$0.01 & 0.51$\pm$0.02 & 0.344$\pm$0.017 &\\    
%14                & 0.40$\pm$0.02 & 0.37$\pm$0.03 & 0.271$\pm$0.019 &    
% One&Two\tablenote{footnote.}&Three&Four\\ % Place \tablenote{}
\end{tabular}
\end{table}
\end{minipage}

\vspace{-0.4cm}
%%%%%%%%%%%%%%%%%%%%%%%%%%%%%%%%%%%%%%%%%%%%%%%%%%%%%%%%%%%%%%%%%

\subsection{Dielectric loss with fixed frequency}
%We now discuss how the dynamics of the $\alpha$-process change with  
%decreasing $T_{\rm g}$ resulting from decreasing $d$. 
Figure 4 shows the reduced dielectric loss 
$\tilde\epsilon''/\tilde\epsilon''_{\rm max}$ as 
a function of temperature 
at 100Hz in a-PS samples of thickness 
9nm, 26nm and 105nm with $M_{\rm w}$=2.8$\times$10$^5$.
Here, the reduced dielectric loss is 
defined by $\tilde\epsilon ''/\tilde\epsilon ''_{\rm
max}=(C''(T)-C''(T_0))/(C''_{\rm max}(T_{\alpha})-C''(T_0))$, where $C''$ is the 
imaginary part of the complex capacitance, $T_0$ is a standard temperature
(in this case, $T_0$ is room temperature), and $C''_{\rm max}$ 
is the peak value of $C''(T)$ due to the $\alpha$-process. 
Above $T_{\rm g}$ the dielectric loss
$\tilde\epsilon''/\tilde\epsilon''_{\rm max}$ for a given frequency 
displays an anomalous increase with temperature due to the
$\alpha$-process, and it possesses a maximum at the temperature 
$T_{\alpha}$. 
The value of $T_{\alpha}$ and the width of the $\alpha$-peak,  
$\Delta T_{\alpha}$, also depend on $d$, as shown in Fig.4. Here, 
$\Delta T_{\alpha}$ is defined as the temperature difference 
between $T_{\alpha}$ and the lower temperature 
at which $\tilde\epsilon ''$ is half its peak value. 
As shown in Fig.5(b), the width $\Delta T_{\alpha}$
 begins to increase at about 100 nm and continues to increase monotonically 
with decreasing $d$. The $d$ dependence of 
$\Delta T_{\alpha}$ can be expressed as 
\begin{eqnarray}\label{Del_alpha}
\Delta T_{\alpha}(d)=\Delta T_{\alpha}^{\infty}
 \left(1+\frac{a'}{d}\right), 
\end{eqnarray}
where $a'$=6.9$\pm$0.8 nm and 
$\Delta T_{\alpha}^{\infty}$=10.3$\pm$0.4 K. Comparing the $d$-dependence of 
$\Delta T_{\alpha}$ with that of $T_{\rm g}$ (Fig.2), we find that
the decrease of $T_{\rm g}$ is directly correlated with the 
broadening of the $\alpha$-peak as 
$\delta(T_{\rm g}(d))/T_{\rm g}^{\infty}$=
$-C_1 %-6.0\times 10^{-2} 
\delta(\Delta T_{\alpha}(d))/
\Delta T_{\alpha}^{\infty}$, 
where $\delta T_{\rm g}(d)$=$T_{\rm g}(d)-T_{\rm g}^{\infty}$, 
$\delta(\Delta T_{\alpha}(d))$=$\Delta T_{\alpha}(d)-\Delta 
T_{\alpha}^{\infty}$, 
and $C_1$ is a constant 
(4.8$\times$10$^{-2}$$-$6.5$\times$10$^{-2}$).
In other words, it can be concluded that the broadening of 
the distribution of relaxation times 
for the $\alpha$-process is closely correlated to the decrease 
of $T_{\rm g}$. 

Contrastingly, Fig.5(a) shows that $T_{\alpha}$ remains 
almost {\it constant} as $d$ is decreased, down to the critical
thickness $d_{\rm c}$, at which point it begins to {\it decrease linearly}
with decreasing $d$. Therefore, $T_{\alpha}$ is given as follows:
\begin{eqnarray}\label{Ta_d}
T_{\alpha}(d)= \left\{ \begin{array}{c@{\quad:\quad}l} 
T_{\alpha}^{\infty} & d > d_{\rm c}\\ 
T_{\alpha}^{\infty}\left(1+\frac{d-d_{\rm c}}{\zeta}\right) & d < d_{\rm c},
\end{array}\right. 
\end{eqnarray}
where $T_{\alpha}^{\infty}$ and $\zeta$ are constants.
The functional form of $T_{\alpha}$ with respect to $d$ is independent
of $M_{\rm w}$, because Eq.(\ref{Ta_d}) can well reproduce experimental 
values of $T_{\alpha}$ for two different molecular weights 
$M_{\rm w}$=2.8$\times$10$^5$ and 1.8$\times$10$^6$. The parameters 
$d_{\rm c}$ and $\zeta$ show a distinct molecular weight
dependence as shown in Table II. The $M_{\rm w}$  and $d$ dependences 
of $T_{\alpha}$ are quite different from those of $T_{\rm g}$ and 
$\Delta T_{\alpha}$ found in the present and previous  
measurements on supported PS films~\cite{Keddie1,Forrest1a}. They  
are similar to those of $T_{\rm g}$ for freely standing films of 
a-PS~\cite{Forrest1}.

The values of $d_{\rm c}$ listed in Table II are 
%clearly depend on the molecular weight ($M_{\rm w}$) of a-PS: 
$d_{\rm c}$=11 nm for $M_{\rm w}$=2.8$\times$10$^5$ and
$d_{\rm c}$=20$\sim$23 nm for $M_{\rm w}$=1.8$\times$10$^6$. 
These values seem to be related to the radius of gyration
of the bulk polymer coil ($R_{\rm g}$=0.028$\times$$\sqrt{M}$
(nm)~\cite{DeMaggio}):
$R_{\rm g}$=15 nm for $M_{\rm w}$=2.8$\times$10$^5$ and 38 nm for
 $M_{\rm w}$=1.8$\times$10$^6$. 
Furthermore, if we assume that $d_{\rm c}$ and $\zeta$ can be scaled by
the functional form of $d_{\rm c}\sim M^{\epsilon}$ and $\zeta\sim
M^{\gamma}$, where $M$ is the molecular weight of polymers, the values of 
$\epsilon$ and $\gamma$ can be estimated 
as follows: $\epsilon$=0.38$\pm$0.10 and $\gamma$= 0.51$\pm$ 0.08. These 
values are nearly 
equal or similar to the exponent of radius of
gyration for Gaussian chains ($\nu =0.5$). In the 
molten state, polymer chains can be regarded as Gaussian
chains. This result suggests that the length scale such as the radius 
of gyration of the bulk polymer may control the drastic change of 
$T_{\alpha}$ near and below $d_{\rm c}$. In other words, the deformation 
of random coils of polymer chains confined in thin films may cause 
the observed decrease in $T_{\alpha}$ with decreasing film thickness. 
As discussed later, one of the possible origin of this deformation is
the competition between the liquid-like layer and the dead layer.
The investigation of
$d$ dependence of $T_{\alpha}$ for various $M_{\rm w}$ samples will 
reveal the detailed mechanism of drastic decrease in $T_{\alpha}$ with 
decreasing $d$.

Here, it should be noted that the length scale $a'$ included in the 
expression of $\Delta T_{\alpha}$ as a function of $1/d$ is much larger 
than $a$ in $T_{\rm g}$: $a'/a=15.3$. This can be explained 
in the following way. It is assumed that 
the shape of the loss peak due to the $\alpha$-process in the plot of 
$\epsilon ''$ vs. $T$ can be expressed by the same function of $T_{\alpha}$
and $\Delta T_{\alpha}$ for any film thickness $d$ and that $T_{\rm g}$ 
can be regarded as the
temperature at which dielectric loss begins to increase due to the 
$\alpha$-process.
From this assumption, the following relation can be obtained:
\begin{eqnarray}\label{Ta_Tg}
T_{\alpha}(d)-T_{\rm g}(d)= A\times\Delta T_{\alpha}(d),
\end{eqnarray}
where $A$ is a constant. Taking into account that for $d>d_{\rm c}$, 
$T_{\alpha}(d)=T_{\alpha}^{\infty}$, and substituting Eqs.(\ref{Tg_d})
and (\ref{Del_alpha}) into Eq.(\ref{Ta_Tg}), we obtain the relations
$T_{\alpha}^{\infty}-T_{\rm g}^{\infty}=A\Delta T_{\alpha}^{\infty}$ and 
$T_{\rm g}^{\infty}a=A\Delta T_{\alpha}^{\infty}a'$. Hence, the value 
of $a'/a$ is expressed by 
\begin{eqnarray}\label{a1-a}
\frac{a'}{a}=\frac{T_{\rm g}^{\infty}}{T_{\alpha}^{\infty}-
T_{\rm g}^{\infty}}.
\end{eqnarray}
Using Eq.(\ref{a1-a}) with the observed values of $T_{\rm g}^{\infty}$ 
and $T_{\alpha}^{\infty}$ we obtain $a'/a\approx 17$ for $f$=100Hz and 
12 for $f$=1kHz. Although the errors in $\Delta T_{\alpha}$ prevent
us from obtaining the frequency dependence of $a'$, the values
of $a'/a$ evaluated using the above assumption agree well with those
found in the present measurements.

%%%%%%%%%%% FIG.6 %%%%%%%%%%%
\hspace*{-0.4cm}
%\vspace{-0.5cm}
\begin{minipage}{8.5cm}
\begin{figure}
\epsfxsize=7.8cm %\epsfysize=10cm
%\epsfxsize=14cm %\epsfysize=10cm
%\vspace*{3cm}
%\centerline{
\vspace*{-0.4cm}
\epsfbox{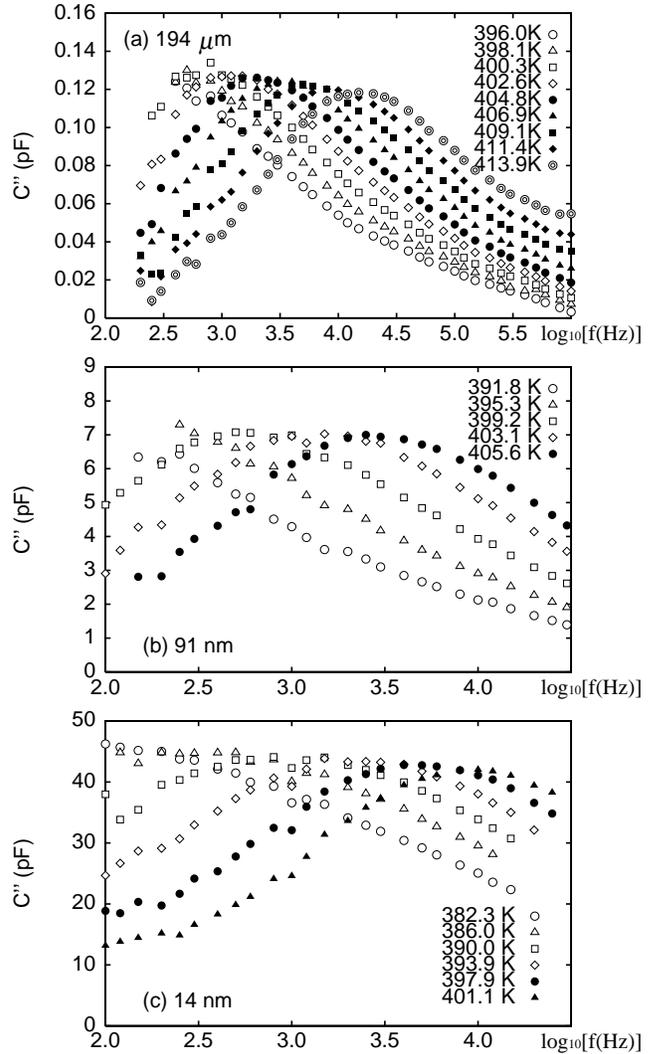}
%}
\vspace{-2.3cm}
\caption{Frequency dependence of the imaginary part of the complex capacitance
for a-PS thin films with various film thicknesses 
($M_{\rm w}$=1.8$\times$10$^6$): (a) $d$=194 $\mu$m (bulk sample); (b) 
$d$=91 nm; (c) $d$=14 nm.
}
\label{fig:fig6}
\end{figure}

\vspace{0.1cm}
\end{minipage}
%%%%%%%%%%%%%%%%%%%%%%%%%%%%%

Recently, Forrest {\it et al.} obtained $\alpha$-relaxation 
data with a characteristic time scale 
$\langle\tau\rangle$$\sim$2$\times$10$^{-4}$s using a quartz crystal 
microbalance technique applied to supported PS films covered with SiC 
particles~\cite{Forrest2}. 
It was reported that the small dissipation peak $T_{\rm max}$, which 
corresponds to $T_{\alpha}$ in this paper, exhibits the same $d$ dependence 
as $T_{\rm g}$ when the values of $T_{\rm g}$ are shifted by 20K.
The $d$ dependence of $T_{\rm max}$ found from their measurements 
seems to be different from that found from the present  measurements.
The values of $T_{\rm g}$ used for comparison with $T_{\rm max}$
were observed by ellipsometry for PS films supported on a
hydrogen-passivated Si substrate with a free surface~\cite{Keddie3}.
Hence, the comparison between $T_{\rm g}$ and $T_{\rm max}$ in
Ref.\cite{Forrest2} was carried out for samples with different molecular weight 
and of different geometries (with and without a free surface) by using 
%%%%%%%%%%% FIG.7 %%%%%%%%%%%
\hspace*{-0.3cm}
%\vspace{-0.5cm}
\begin{minipage}{8.5cm}
\begin{figure}
\epsfxsize=8.8cm %\epsfysize=10cm
%\vspace*{8cm}
\centerline{
\hspace*{-0.5cm}\epsfbox{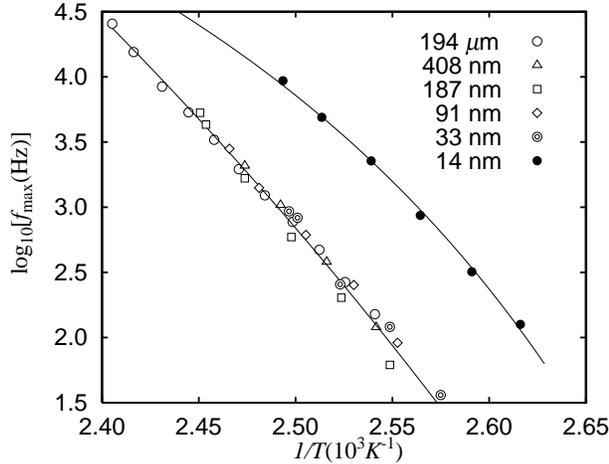}
%\epsfbox{T_alpha_6.eps}
}
\vspace{0.4cm}
\caption{Peak frequency of dielectric loss due to the $\alpha$-process
as a function of the inverse of temperature for thin films 
of a-PS with various film thicknesses ($M_{\rm w}$=1.8$\times$10$^6$). 
Solid curves were obtained by fitting the data to the VFT equation, 
$f_{\rm max}=f_0\exp (-U/(T-T_V))$, where $f_0$, $U$, and $T_V$ are 
constants.
}
\label{fig:fig7}
\end{figure}

\vspace{0.1cm}
\end{minipage}
%%%%%%%%%%%%%%%%%%%%%%%%%%%%%
different experimental techniques, while a direct comparison was 
carried out in our study by using simultaneous 
measurements of $T_{\rm g}$ and $T_{\alpha}$ for the same sample.
The results obtained in Ref.\cite{Forrest2} may have been affected by a small
difference in experimental conditions for the two different measurements of
$T_{\rm g}$ and $T_{\rm max}$. Furthermore, $T_{\rm max}$ plotted 
in the inset of Fig.2 of Ref.\cite{Forrest2} can also be fitted by 
the equation proposed for $T_{\alpha}$ in this paper 
(See Eq.(\ref{Ta_d})), and the critical thickness $d_{\rm c}$
is found to be 35 nm.
%Since only 6 data points of $T_{\rm max}$ are available, 
%more data points are required to judge which functinal form with respect 
%to $d$ can well reproduce the observed $T_{\rm max}$.

%This may come from the difference between dielectric relaxation dynamics 
%and mechanical relaxation dynamics.

\subsection{Dielectric relaxation behavior of thin films}

Here, we give the results for the imaginary part of the complex capacitance 
(dielectric loss) as a function of the frequency to facilitate
discussion of the 
dynamics of thin films of a-PS with various film thicknesses between 14 nm
and 194 $\mu$m.  Figure 6 displays the dielectric loss vs. frequency at various 
temperatures above $T_{\rm g}$ for a-PS ($M_{\rm w}$=1.8$\times$10$^6$)
with film thicknesses of (a) 194 $\mu$m (bulk sample), (b) 91 nm 
and (c) 14 nm. The peak 
in Fig.6 corresponds to that due to the $\alpha$-process. 
It it found that the relaxation behavior of the $\alpha$-process changes
with temperature and thickness. The 
peak frequency shifts to the higher frequency side as the temperature
increases. In Fig.7 the peak frequency $f_{\rm max}$, which corresponds to 
the inverse of the relaxation time $\tau_{\alpha}$ of the $\alpha$-process,  
is plotted as a function of the inverse temperature. 
It is found that the values of $\tau_{\alpha}$ for the films with
thickness from 33 nm to
194 $\mu$m fall on {\it the same curve}, which can be described by the 
Vogel-Fulcher-Tammann (VFT) equation.
Here it should be noted that the values of  
$T_{\rm g}$ for the films with thickness of 33 nm and 91 nm are 
%%%%%%%%%%% FIG.8 %%%%%%%%%%%
%\hspace*{-0.2cm}
%\vspace{-0.5cm}
\begin{minipage}{8.5cm}
\begin{figure}
\epsfxsize=7.8cm %\epsfysize=10cm
%\vspace*{3cm}
%\centerline{
\hspace*{-1.2cm}
\vspace*{-1.5cm}

\epsfbox{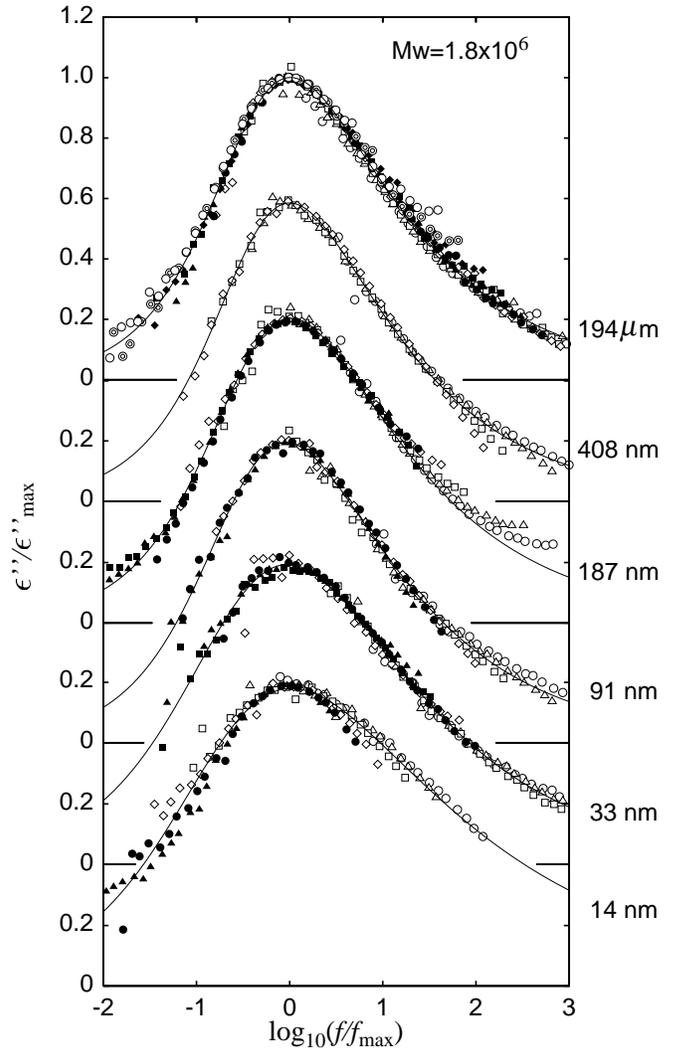}
%}
\vspace{3.9cm}
\caption{Dependence of the normalized dielectric loss on the logarithm
of the normalized frequency ($M_{\rm w}$=1.8$\times$10$^6$). The two 
axes are normalized with respect to the peak position due to 
the $\alpha$-process, corresponding to $\epsilon_{\rm max}''$ and $f_{\rm max}$. 
The numbers given in the right margin are the film thicknesses. 
The solid curves for each film with various thicknesses were obtained 
by fitting the data to the HN equation. The parameter values resulting 
in the best fit are given 
in Table II. For $d$=194 $\mu$m, 91 nm and 14 nm, the symbols 
are the same as in Fig.6. For $d$=408 nm, the symbol $\circ$ corresponds
to 393.5K, $\triangle$ to 397.5K, $\Box$ to 401.3K, and $\Diamond$ to 
404.3K. For $d$=187 nm, the symbol $\circ$ corresponds to 388.5K, 
$\triangle$ to 392.4K, $\Box$ to 396.3K, $\Diamond$ to 400.4K, 
$\bullet$ to 404.3K, \protect\ftriangle{0.5mm} to 407.6K, and 
\protect\fhako{0.5mm}
to 408.1K. For $d$=33 nm, the symbol $\circ$ corresponds to 380.5K, 
$\triangle$ to 384.5K, $\Box$ to 388.4K, $\Diamond$ to 392.4K, 
$\bullet$ to 396.4K, \protect\ftriangle{0.5mm} to 399.9K, and 
\protect\fhako{0.5mm} to 400.6K.
}
\label{fig:fig8}
\end{figure}

\vspace{0.1cm}
\end{minipage}
%%%%%%%%%%%%%%%%%%%%%%%%%%%%%
smaller
than those of thicker films, although 
$\tau_{\alpha}$ remains constant. As the thickness decreases further,
down to 14nm, $\tau_{\alpha}$ becomes much shorter than that for  
thicker films. It follows from this result that the relaxation time
$\tau_{\alpha}$ of the $\alpha$-process 
remains constant down to a critical thickness, below which 
it begins to decrease. 
This result 
%%%%%%%%%%% FIG.9 %%%%%%%%%%%
\hspace*{-0.2cm}
%\vspace{-0.5cm}
\begin{minipage}{8.5cm}
\begin{figure}
\epsfxsize=9cm %\epsfysize=10cm
%\vspace*{6cm}
\centerline{
\hspace*{-0.5cm}\epsfbox{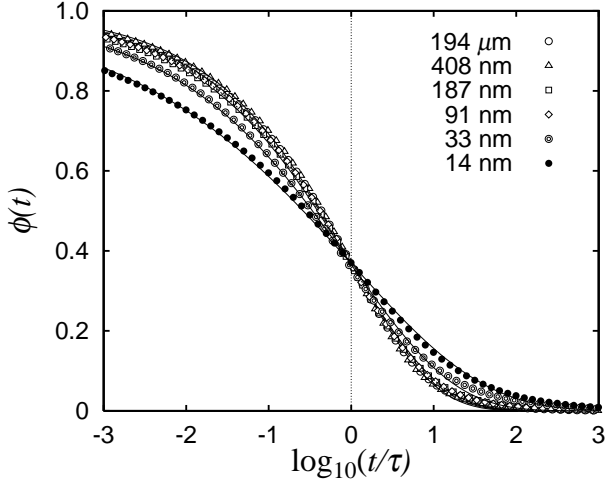}
}
\vspace{0.4cm}
\caption{The relaxation function $\phi$ as a function of the logarithm 
of the reduced time $t/\tau_{\mbox{\protect\tiny KWW}}$ for a-PS thin films 
with various thicknesses.
The relaxation functions plotted by the 6 different symbols were calculated 
using Eq.(\protect\ref{HNeq}) and Eq.(\protect\ref{relax_fn1}) with 
the best-fit parameters $\alpha_{\rm HN}$ 
and $\beta_{\rm HN}$ listed in Table II for each thickness. The solid
curves were obtained by fitting the relaxation functions to the KWW
equation.
}
\label{fig:fig9}
\end{figure}

\vspace{0.1cm}
\end{minipage}
%%%%%%%%%%%%%%%%%%%%%%%%%%%%%
%%%%%%%%%%%%%% Table III %%%%%%%%%%%%%
%\hspace*{-0.2cm}
\begin{minipage}{8.5cm}
\begin{table}
%\vspace*{5cm}
\caption{The values of $\protect\alpha_{\rm HN}$, $\protect\beta_{\rm HN}$ 
and  $\protect\beta_{\mbox{\protect\tiny KWW}}$ resulting in the best fit for 
thin films of a-PS with $M_{\rm w}$=1.8$\times$10$^6$.}
\begin{tabular}{cccc} % In second brace, l = left, r = right,
% c = centered and d = decimal justification.
$d$ (nm) & $\alpha_{\rm HN}$ & $\beta_{\rm HN}$ & $\beta_{\mbox{\tiny KWW}}$ 
\\\tableline 
194$\times$10$^3$ & 0.22$\pm$0.01 & 0.46$\pm$0.01 & 0.419$\pm$0.012\\     
408               & 0.22$\pm$0.01 & 0.48$\pm$0.02 & 0.435$\pm$0.017\\   
187               & 0.25$\pm$0.01 & 0.46$\pm$0.03 & 0.399$\pm$0.021\\   
91                & 0.25$\pm$0.01 & 0.47$\pm$0.01 & 0.406$\pm$0.014\\    
33                & 0.38$\pm$0.01 & 0.51$\pm$0.02 & 0.344$\pm$0.017\\    
14                & 0.40$\pm$0.02 & 0.37$\pm$0.03 & 0.271$\pm$0.019     
% One&Two\tablenote{footnote.}&Three&Four\\ % Place \tablenote{}
\end{tabular}
\end{table}
\end{minipage}
%%%%%%%%%%%%%%%%%%%%%%%%%%%%%%%%%%%%%%%%%%%%%%%%%%%%%%%%%%%%%%%%%%%%5
is consistent with that extracted from the
experimental observations of dielectric loss with fixed frequency
discussed in Sec.IV.A.

In order to discuss the relaxation behavior of the $\alpha$-process,
the profiles in Fig.6 are replotted by scaling them with respect to 
peak positions and peak heights. Figure 8 shows that 
profiles of dielectric loss vs. frequency can be reduced in this way to a 
single master curve over the 
temperature range above $T_{\rm g}$ described in the figure captions. 
It it clearly found that peak profiles become broader 
as the film thickness decreases. The master curve for each film with
different thickness can be fitted by using the Havriliak-Negami 
equation~\cite{HN},
\begin{eqnarray}\label{HNeq}
\epsilon ''(\omega)= \Im \frac{\Delta\epsilon}{[1+(i\omega\tau_0)^{1-\alpha_{HN}}]^{\beta_{HN}}},
%\nonumber
\end{eqnarray}
where $\Delta\epsilon$ is the dielectric strength, $\omega$ is angular 
frequency (= $2\pi f$), $\tau_0$ is the apparent relaxation time,
and $\alpha_{HN}$ and $\beta_{HN}$ are shape
parameters. The values of $\alpha_{\rm HN}$ and $\beta_{\rm HN}$
resulting in the best fit
are given in Table III. The 
%%%%%%%%%%% FIG.10 %%%%%%%%%%%
\hspace*{-0.2cm}
%\vspace{-0.5cm}
\begin{minipage}{8.5cm}
\begin{figure}
\epsfxsize=9.0cm %\epsfysize=10cm
%\vspace*{8cm}
\centerline{
\hspace*{-0.5cm}\epsfbox{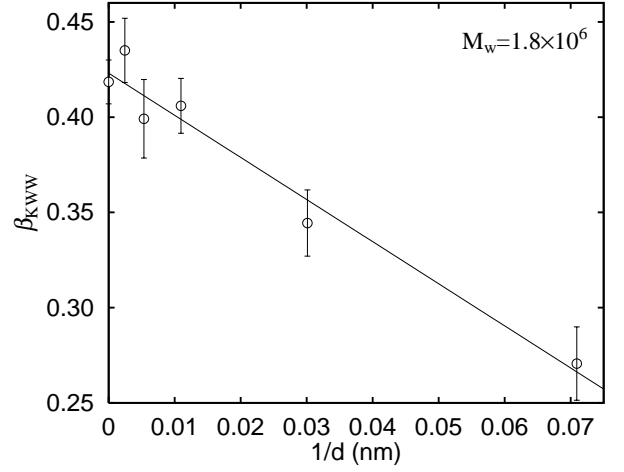}
}
\vspace{0.3cm}
\caption{The KWW exponent $\beta_{\mbox{\protect\tiny KWW}}$ as a function of the 
inverse of the thickness $d$ ($M_{\rm w}$=1.8$\times$10$^6$). 
The values $\beta_{\mbox{\protect\tiny KWW}}$ were obtained by fitting 
the relaxation function to the KWW equation. 
The solid line was plotted using Eq.(\protect\ref{BetaKWW}).
}
\label{fig:fig10}
\end{figure}

\vspace{0.1cm}
\end{minipage}
%%%%%%%%%%%%%%%%%%%%%%%%%%%%%
solid curves given in Fig.8 are calculated 
using the HN equation. 

Here, the data in the frequency domain are converted into those in time
domain. Using the equation
\begin{eqnarray}\label{relax_fn1}
\phi(t)&=&
\frac{2}{\pi} \int_0^{\infty} \frac{\epsilon''(\omega)}{\Delta\epsilon} 
\cos\omega t \frac{d\omega}{dt}, %\nonumber
\end{eqnarray}
the relaxation function $\phi(t)$ can be calculated via the HN equation 
with the best-fit parameters for thin films of various thicknesses, as 
shown in Fig.9. 
The shape of the relaxation function changes with film thickness. 
As seen in Fig.9, the relaxation function thus obtained can be fitted 
quite well by the Kahlrausch-Williams-Watts (KWW) equation
\begin{eqnarray}\label{KWW}
\phi (t)= \exp\left[-\left(\frac{t}{\tau_{\mbox{\tiny KWW}}}\right)^{\beta_{\mbox{\tiny KWW}}}\right],
%\nonumber
\end{eqnarray}
for any thickness. It is also found that the relaxation behavior becomes 
slower as the film thickness decreases. 

Figure 10 displays the exponent $\beta_{\mbox{\tiny KWW}}$ 
as a function of the inverse of the film thickness.
It it found that $\beta_{\mbox{\tiny KWW}}$ decreases 
from 0.42 to 0.27 as the thickness changes from 194 $\mu$m to
14 nm. The functional form of $\beta_{\mbox{\tiny KWW}}$ with respect 
to the inverse of the thickness is found to be linear:
\begin{eqnarray}\label{BetaKWW}
\beta_{\mbox{\tiny KWW}}=\beta_{\mbox{\tiny KWW}}^{\infty}\left(1-\frac{a''}{d}\right),
\end{eqnarray}
where $\beta_{\mbox{\tiny KWW}}^{\infty}$=0.423$\pm$0.006 and
$a''$=(5.2$\pm$0.4) nm. 
The value of $\beta_{\mbox{\tiny KWW}}$ is a measure of the distribution 
of relaxation times $\tau_{\alpha}$ of the $\alpha$-process; $i.e.$, 
the distribution becomes broader as $\beta_{\mbox{\tiny KWW}}$ becomes smaller.
Therefore, as the thickness decreases, the distribution of the relaxation 
times $\tau_{\alpha}$ becomes broader according to Eq.(\ref{BetaKWW}).
%%%%%%%%%%% FIG.11 %%%%%%%%%%%
\hspace*{-0.2cm}
%\vspace{-0.5cm}
\begin{minipage}{8.5cm}
\begin{figure}
\epsfxsize=8.8cm  %\epsfysize=11.2cm
%\vspace*{4cm}
\centerline{
\hspace*{-0.5cm}\epsfbox{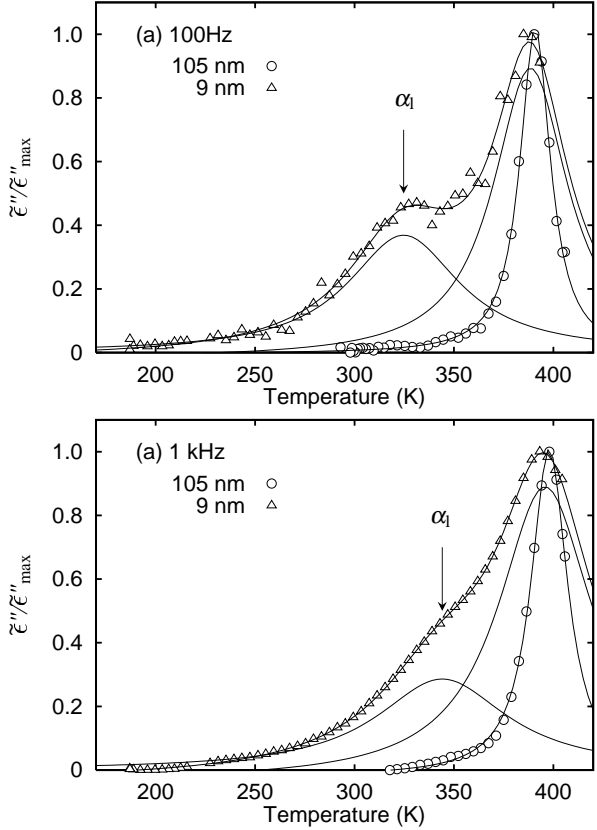}
}
\vspace{-2.3cm}
\caption{Reduced dielectric loss as a function of temperature during
the cooling process for a-PS with film thicknesses of 105 nm and 9 nm 
($M_{\rm w}$=2.8$\times$10$^5$): (a) $f$=100Hz and (b) $f$=1kHz.
The solid curves for the $\alpha_{\rm l}$-process were obtained by 
fitting the data to the equation 	   
$\tilde\epsilon''=\sum_{j=\alpha,\alpha_{\rm l}}
\tilde\epsilon''_{{\rm max}, j}/(1+((T-T_j)/\Delta T_j)^2)$.
}
\label{fig:fig11}
\end{figure}

\vspace{0.1cm}
\end{minipage}
%%%%%%%%%%%%%%%%%%%%%%%%%%%%%
The characteristic length scale $a''$ obtained in this analysis of
$\beta_{\mbox{\tiny KWW}}$ is consistent with the value $a'$ obtained for 
$\Delta T_{\alpha}$. 
Furthermore, the relative change of 
$\beta_{\mbox{\tiny KWW}}$ measured with respect to that of the bulk sample, 
$\delta\beta_{\mbox{\tiny KWW}}(d)$
=$\beta_{\mbox{\tiny KWW}}^{\infty}-\beta_{\mbox{\tiny KWW}}(d)$, 
is directly related to the relative change of $T_{\rm g}$ as follows:
\begin{eqnarray}
\frac{\delta T_{\rm g}(d)}{T_{\rm g}^{\infty}} 
= 9.6\times 10^{-2} \times \frac{\delta\beta_{\mbox{\tiny KWW}}(d)}
{\beta_{\mbox{\tiny KWW}}^{\infty}}.
\end{eqnarray}

The values of $\beta_{\mbox{\tiny KWW}}$ for freely standing 
films of a-PS have been evaluated using photon correlation spectroscopy
and found to be indistinguishable from those of bulk PS~\cite{Forrest2}. 
On the other hand, a decrease in $\beta_{\mbox{\tiny KWW}}$ has been  
observed for a copolymer thin film supported on quartz~\cite{Hall}, as found 
in the present measurements on PS supported on a glass substrate.
We thus conclude that there is a large difference in the dynamics of freely
standing films and supported thin films. In the case of thin films supported 
on substrate, it is easily understood that the existence of the substrate
may cause a broadening of the distribution of $\alpha$-relaxation times,
because the dynamics of polymer chains should depend on the 
distance from the substrate near the boundary.

Although only the results for the dielectric loss in thin films 
of a-PS with $M_{\rm w}$=1.8$\times$10$^6$ are given in IV.B,  
the results obtained for thin films with $M_{\rm w}$=2.8$\times$10$^5$ 
are consistent with these.

\section{Discussion and Summary}
In the case of thin polymer films supported on substrate, not only the
 surface effects but also the interactions between the substrate and
 films strongly affect the dynamics and the glass transition of the 
thin films. We introduced a {\it three-layer model} in order to 
explain such surface and interfacial effects in Sec.III.B, 
following Ref.\cite{DeMaggio}. 
In this model it is assumed that within a thin film there are a liquid-like 
layer and a dead layer in addition to a bulk-like layer. 

According to this model, two different $\alpha$-processes should exist 
corresponding to the liquid-like layer and bulk-like layer.
(If the $\alpha$-process exists in the dead layer, it should exist only far above
$T_{\rm g}$ of the bulk sample, $i.e.$, beyond the experimentally accessible 
temperature range.)
We now are performing dielectric measurements to investigate the 
dynamical properties of a-PS thin films over a wider temperature
range. Although the measurements are still in progress, direct 
evidence for the existence of different processes in thin films 
of a-PS is displayed in Fig.11.
Figure 11 shows the normalized dielectric loss as a function of 
temperature for a-PS thin films with thicknesses of 105 nm and 9 nm and 
$M_{\rm w}$=2.8$\times$10$^5$. In this case, the standard temperature 
$T_0$ is set to 180 K. For films with $d$=105 nm and 9 nm there the
$\alpha$-process exists around 390 K for $f$=100Hz, as discussed in Sec.IV, 
while for films with $d$=9 nm there another 
$\alpha$-process exists at lower temperature. We refer to the latter as
the $\alpha_{\rm l}$-process (see arrows in the figure).
This peak in the dielectric loss due to the $\alpha_{\rm l}$-process
shifts to the higher temperature side as the frequency of the applied 
electric field changes from 
100Hz to 1kHz. This suggests that the dynamical behavior of the peak
is similar to that of the $\alpha$-process.
Since within the model the thickness of the liquid-like layer is assumed to be 
constant (independent of the thickness $d$) the contribution 
of the liquid-like layer should become more appreciable as the thickness 
decreases.
Therefore, it is reasonable to attribute the loss peak for the $\alpha_{\rm
l}$-process to the liquid-like layer. 
In other words, the existence of the $\alpha_{\rm l}$-process can 
be regarded as experimental evidence of the existence of a liquid-like 
layer within a thin film.

Since the $\alpha$-process and the $\alpha_{\rm l}$-process can be 
attributed to the segmental motion of polymer chains in the bulk-like 
layer and liquid-like layer (surface layer), respectively, we can expect 
that the characteristic time of the $\alpha_{\rm l}$-process is smaller 
than that
of the $\alpha$-process at a given temperature, because polymer chains
involved in the $\alpha_{\rm l}$-process has a higher mobility 
than those in the $\alpha$-process of bulk-like layer. 
At higher temperatures %(far above the glass transition temperature) 
the segmental motions of the surface layer is expected to become less
different from those of the bulk-like layer.
Therefore, $T_{\alpha_l}$ should be smaller than $T_{\alpha}$
at a given frequency and the difference between $T_{\alpha}$ and 
$T_{\alpha_{\rm l}}$ should become smaller with increasing frequency.
Here, $T_{\alpha_{\rm l}}$ is the temperature at which the dielectric loss
possesses a maximum due to the $\alpha_{\rm l}$-process at a given 
frequency.
Because the chain end concentration in the surface layer is higher 
than that in the bulk-like layer, $T_{\alpha_{\rm l}}$ can be expected 
to have a stronger molecular-weight dependence than $T_{\alpha}$ does.

If there are liquid-like and dead layers in addition 
to a bulk-like layer, it is plausible that there are boundary layers 
between the bulk-like layer and the dead layer and between the bulk-like
layer and the liquid-like layer. The 
existence of such boundary layers causes the broadening of the 
distribution of the $\alpha$-relaxation time of the bulk-like layer, 
and this broadening is enhanced as the thickness decreases. 
%The $d$-dependence 
%of $\alpha_{\rm n}$ observed in our measurements also supports the 
%model~\cite{Fukao1}. 
%On the basis of this model, the existence of mobile and immobile 
%layers with constant layer thicknesses broadens the distribution of
%relaxation times of the $\alpha$-process with decreasing $d$, $i.e.$, 
%$\Delta T_{\alpha}$ is increased. 
Because $T_{\rm g}$ can be regarded 
as the temperature at which the anomalous increase in $\epsilon''$ 
begins, $T_{\rm g}$ decreases as the distribution of $\tau_{\alpha}$
broadens; $i.e.$, $\Delta T_{\alpha}$ increases or 
$\beta_{\mbox{\tiny KWW}}$ decreases.

Since the liquid-like and dead layers cause the distribution of $\tau_{\alpha}$ 
to broaden toward the shorter time side and the longer time side, respectively, 
the relaxation time $\tau_{\alpha}$ can remain constant as long 
as the two contributions cancel each other.
In the present measurements, we measured the peak frequencies $f_{\rm max}$ in
the dielectric loss due to the $\alpha$-process, which are equal to the 
inverse of $\tau_{\alpha}$, and it is found that $f_{\rm max}$ at a given
temperature is almost independent of the thickness down to the critical 
thickness $d_{\rm c}$.
%If the average relaxation time of $\alpha$-process can be evaluated,
%the dependence of averaged $\tau_{\alpha}$ on thickness may be
%observed even above $d_{\rm c}$.
 
Below the critical thickness, there is no $\alpha$-process 
with dynamical properties observed in the bulk sample. In this case, the dynamical 
properties of the thin film are determined by the competition between
the liquid-like layer and the dead layer within the thin film.
The values of $\tau_{\alpha}$ can no longer remain constant, 
but it decreases or increases, depending on whether contributions from 
the liquid-like layer are stronger or weaker than those from the dead layer.
In the present case, $\tau_{\alpha}$ and $T_{\alpha}$ decrease
drastically for $d<d_{\rm c}$. This behavior can be accounted for by assuming 
that contributions from the liquid-like layer are much stronger than those 
from the dead layer.

Here, it should be noted that the observed value of $d_{\rm c}$ is
larger than the sum of thickness of the liquid-like layer and the dead
layer, $\xi+\delta$: $d_{\rm c}$=11$\sim$23 nm, while $\xi+\delta\approx 10$ nm.
From this result it follows that for $d$=$d_{\rm c}$, the bulk-like layer 
still exists, but the relaxation time for the $\alpha$-process of 
the bulk-like layer is different from that of the bulk sample. 
This suggests the possibility  that at $d_{\rm c}$ the thickness 
of the bulk-like 
layer ($d-\xi-\delta$) is comparable to the characteristic 
length scale $\xi_{\mbox{\tiny CRR}}$ of the $\alpha$-process.
Therefore, the value of $\xi_{\mbox{\tiny CRR}}$
can be estimated as 1$\sim$13 nm, if there is assumed to exist   
any characteristic length scale of the $\alpha$-process at all.  
Because $d_{\rm c}$ depends on the 
molecular weight of a-PS, $\xi_{\mbox{\tiny CRR}}$ may also 
depend on the molecular weight. However, more precise measurements 
of $\delta$ and $\xi$ showing whether these values depend on the
molecular weight are necessary to elucidate the molecular weight 
dependence of $\xi_{\mbox{\tiny CRR}}$.

%On the other hand, the temperature $T_{\alpha}$ remains constant,
%because $T_{\alpha}$ is the temperature at which $\epsilon''$ has 
%a maximum. 
%(The $\alpha$-relaxation time averaged over all the
% distributions may change, even if $T_{\alpha}$ does not change.) 
%If $d$ reaches a critical thickness $d_{\rm c}$, the thickness of 
%the bulk-like layer becomes comparable to the characteristic length 
%scale of the $\alpha$-process and, as a result, the dynamics 
%change drastically. For $d$$<$$d_{\rm c}$, $T_{\alpha}$ decreases or
% increases depending on whether contributions from the mobile layer 
%are stronger than those from the immobile layer.

In this paper, four different length scales, $a$, $a'$, $a''$ and $d_{\rm c}$, 
were extracted from the dielectric measurements. 
From the above discussion, $d_{\rm c}$ is expected to be related to 
the characteristic length scale $\xi_{\mbox{\tiny CRR}}$ 
for the $\alpha$-process of bulk
samples, while the values of $a$, $a'$ and $a''$ are believed to be
related to the 
heterogeneous structure of the mobility within thin films or surface and 
interfacial effects.   
According to the model of Adam and Gibbs, the size of the CRR increases 
as the temperature approaches  $T_{\rm g}$. 
Near the glass transition temperature, the characteristic time 
of the $\alpha$-process is larger than 10$^3$ sec, and only 
the slow modes contribute appreciably to the dielectric 
loss. Hence, a more prominent $d$-dependence of the $\alpha$-process 
can be expected when the dielectric measurements are performed 
for frequencies much smaller than those adopted in the present work. 
We plan to make such measurements in the future.

In this paper, we made dielectric measurement on capped thin films of atactic 
polystyrene supported on an Al deposited glass substrate. The results 
can be summarized as follows:
\begin{enumerate}
\item
The glass transition temperature of thin films of a-PS has 
been determined using the temperature change in the electric capacitance. 
The observed $T_{\rm g}$ is consistent with the results obtained by 
other methods. A decrease in $T_{\rm g}$ with decreasing thickness 
has been confirmed.
\item
The thermal expansion coefficient normal to the film surface 
$\alpha_{\rm n}$ increases with decreasing thickness below the apparent
$T_{\rm g}$, while it decreases with decreasing thickness above $T_{\rm g}$.
The thickness dependence of $\alpha_{\rm n}$ can be described by a
linear function of the inverse of the thickness.
\item 
The $d$ dependence of $T_{\rm g}$ is directly correlated to that 
of the width of the peak due to the $\alpha$-process in the temperature 
domain and also to the distribution of relaxation times of the 
$\alpha$-process. 
\item
The temperature at which dielectric loss is maximal  
due to the $\alpha$-process in the temperature domain and the
$\alpha$-relaxation time obtained by the frequency dependence of the dielectric 
loss remain constant down to the critical thickness $d_{\rm c}$, while
below $d_{\rm c}$ they decreases drastically with decreasing thickness.    
The values of $d_{\rm c}$ have a molecular weight dependence and are
related to the radius of gyration of polymer chains.
\end{enumerate}

\section{Acknowledgments}
This work was partly supported by a Grant-in-Aid from the Ministry 
of Education, Science, Sports and Culture of Japan.

\end{multicols}

\begin{references}
\vspace{-1.5cm}
\bibitem[\dagger]{A} To whom correspondence should be addressed.\\
Electronic address: fukao@phys.h.kyoto-u.ac.jp 
\bibitem{Noncry} ``Proceedings of the 3rd International Discussion 
Meeting on Relaxations in Complex Systems'', J. Non-Cryst. Solids. 
{\bf 235-237} (1998).
\bibitem{Pisa}
``Proceeding of Second Workshop on Non-equilibrium Phenomena in
 Supercooled Fluids, Glasses and Amorphous Materials '',
 J. Phys. Condens. Matter, {\bf 11}(1999)
\bibitem{Ediger1}
M.D. Ediger, C.A. Angell, and S.R. Nagel, J. Phys. Chem. 
{\bf 100}, 13201 (1996), and the references cited therein.
\bibitem{Adam-Gibbs}
G. Adam and J.H. Gibbs, J. Chem. Phys. {\bf 43}, 139 (1965).
\bibitem{Muranaka} 
T. Muranaka and Y. Hiwatari, Phys. Rev. {\bf E51}, R2735, (1995).
\bibitem{Perera} 
D.N. Perera and P. Harrowell, Phys. Rev. {\bf E54}, 1652 (1996).
\bibitem{Yamamoto1}
R. Yamamoto and A. Onuki, J. Phys. Soc. Jpn., {\bf 66}, 2545 (1997), 
Phys. Rev. {\bf E58}, 3515 (1998).
\bibitem{Kob} 
W. Kob, C. Donan, S.J. Plimton, P.H. Poole, and S.C. Glotzer,
 Phys. Rev. Lett. {\bf 79}, 2827 (1997).
\bibitem{Spiess}
K. Schmidt-Rohr and H.W. Spiess, Phys. Rev. Lett. {\bf 66}, 3020 (1991).
\bibitem{Bohmer}
B. Schiener, R. B\"ohmer, A. Loidl, and R.V. Chamberlin, Science, 
{\bf 274}, 752, (1996).
\bibitem{Ediger2}
M.T.~Cicerone and M.D.~Ediger, J. Chem. Phys. {\bf 103}, 5684 (1995).
\bibitem{Schuller}
J. Sch\"uller, Yu. B. Mel'nichenko, R. Richert, and E.W. Fischer, 
Phys. Rev. Lett. {\bf 73}, 2224 (1994).
\bibitem{Kremer1}
M.~Arndt, R.~Stannarius, H.~Groothues, E.~Hempel, and  F.~Kremer, 
Phys. Rev. Lett. {\bf 79}, 2077 (1997).
\bibitem{Barut} 
G. Barut, P. Pissis, R. Pelster, and G. Nimtz, 
Phys. Rev. Lett. {\bf 80}, 3543 (1998); R. Pelster, Phys. Rev. 
{\bf B59}, 9214 (1999).
\bibitem{Keddie1} J.L. Keddie, R.A.L. Jones, and R.A. Cory, Europhys. Lett.,
{\bf 27}, 59 (1994).
\bibitem{Keddie2}
 J.L. Keddie and  R.A.L. Jones, Faraday Discuss. {\bf 98}, 219 (1994).
%positron annihilation
\bibitem{DeMaggio}
G.B. DeMaggio, W.E. Frieze, D.W. Gidley, Ming Zhu, H.A. Hristov, and A.F. Yee, 
Phys. Rev. Lett. {\bf 78}, 1524 (1997).
\bibitem{Forrest1} J.A. Forrest, K. Dalnoki-Veress, J.R. Stevens, and 
J.R. Dutcher, 
Phys. Rev. Lett. {\bf 77}, 2002 (1996).
\bibitem{Forrest1a}
J.A. Forrest, K. Dalnoki-Veress, and J.R. Dutcher, Phys. Rev. 
{\bf E56}, 5705 (1997).
%Thermal expansion, X-ray reflectivity
\bibitem{Wallace}
W.E. Wallace, J.H. van Zanten, and W.L. Wu, Phys. Rev. {\bf E52}, R3329 (1995).
\bibitem{Jerome1}
B. Jerome and J. Commandeur, Nature, {\bf 386}, 589 (1997).
\bibitem{Jerome2}
B. Jerome, J. Phys.: Condens. Matter, {\bf 11}, A189 (1999).
\bibitem{Hall}
D.B. Hall, J.C. Hooker, and J.M. Torkelson, Macromolecules, {\bf 30},
 667 (1997).
\bibitem{Forrest2}
J.A. Forrest, C. Svanberg, K. Revesz, M. Rodahl,
 L.M. Torell, and B. Kasemo,  
Phys. Rev. {\bf E58}, R1226 (1998).
\bibitem{Kajiyama}
T. Kajiyama, K. Tanaka, and A. Takahara, Polymer, {\bf 39}, 4665 (1998).
%Dielectric measurement on LB monolayers.
\bibitem{Fukao1}
K. Fukao and Y. Miyamoto, Europhys. Lett., {\bf 46}, 649 (1999).
\bibitem{Bauer1}
C. Bauer, R. B\"ohmer, S. Moreno-Flores, R. Richert, H. Sillescu,
 D. Neher, Phys. Rev. E, submitted.
\bibitem{Bauer2}
C. Bauer, R. Richert, R. B\"ohmer, J. Non-Cryst. Solids, submitted.
\bibitem{Kremer2}
G. Blum, F. Kremer, T. Jaworek, and G. Wegner, Adv. Mater. {\bf 7}, 1017 
(1995).
%Dielectric measruement on bulk PS.
\bibitem{Yano1}
O.~Yano and Y.~Wada, J. Polym. Sci.: Part A-2, {\bf 9}, 669 (1971).
\bibitem{PolymerHand}
{\it Polymer Handbook}, 3rd ed., edited by J. Brandrup and E.H. Immergut 
 (John Wiley, New York, 1989).
\bibitem{Ballard1}
D.G.H. Ballard, G.D. Wignall, J. Schelten, J. Eur. Polym. J. {\bf 9},
 965 (1973).
\bibitem{Fox1}
T.G. Fox, P.J. Flory, J. Appl. Phys. {\bf 21}, 581 (1950).
\bibitem{Keddie3}
J.L. Keddie, R.A.L. Jones, Israel J. Chem. {\bf 35}, 21 (1995).
\bibitem{Phys1}
{\it Physical Properties of Polymers Handbook}, edited by J.E. Mark,  
pp.337 (AIP Press, New York, 1996).
\bibitem{Tasaka1}
Zhang Xiaomin, S. Tasaka, M. Kondo, and N. Inagaki, Polym. Preprints,
 Jpn. {\bf 48}, E470 (1999). %1029 (1999).
\bibitem{Mayes1}
A.M.~Mayes, Macromolecules, {\bf 27}, 3114 (1994).
\bibitem{Wallace2}
W.E. Wallace, N.C. Beck Tan, W.L. Wu, and S. Satija,  J. Chem. Phys. {\bf 108}, 3798 (1998).
\bibitem{Forrest3}
J.A. Forrest, K. Dalnoki-Veress, and J.R. Dutcher, Phys. Rev. {\bf E58}, 
6109, (1998).
\bibitem{Lee1}
Y.-C. Lee, K.C. Bretz, F.W. Wise, and W. Sachse, Appl. Phys. Lett. 
{\bf 69}, 1692 (1996).
\bibitem{Bretz1}
K.C. Bretz, Y.-C. Lee, F.W. Wise, and W. Sachse,
 Bull. Am. Phys. Soc. {\bf 42}, 648 (1997). 
\bibitem{HN}
S. Havriliak and S. Negami, Polymer {\bf 8}, 161 (1967).
\end{references}
\end{document}